%% file: main.tex
\pdfoutput=1
\documentclass[12pt,a4paper]{article}

\usepackage{ifthen} 
\newboolean{pdflatex}
\setboolean{pdflatex}{true} 

\newboolean{articletitles}
\setboolean{articletitles}{true} 

\newboolean{uprightparticles}
\setboolean{uprightparticles}{false} 

\def\paperauthors{LHCb collaboration} 
\def\paperasciititle{Observation of the Xib->Ximm and study of the Xib->XiPsi(2S) decays} 
\def\papertitle{Observation of the \XibToXimm and study of the \XibToXiPsi decays} 
\def\paperkeywords{{High Energy Physics}, {LHCb}} 
\def\papercopyright{\the\year\ CERN for the benefit of the LHCb collaboration} 
\def\paperlicence{CC BY 4.0 licence}
\def\paperlicenceurl{https://creativecommons.org/licenses/by/4.0/}

\def\LbToLmm{\decay{\Lb}{\Lz \mumu}}

\def\LbToLJPsi{\decay{\Lb}{\Lz \jpsi}}
\def\LbToLPsi{\decay{\Lb}{\Lz \psitwos}}
\def\LbTopKmm{\decay{\Lb}{p\kaon \mumu}}
\def\LbTopKee{\decay{\Lb}{p\kaon \epem}}

\def\XibToXiJPsi{\decay{\Xires^-_b}{\Xires^- \jpsi}}
\def\XibToXiPsi{\decay{\Xires^-_b}{\Xires^- \psitwos}}

\def\XibToXimm{\decay{\Xires^-_b}{\Xires^- \mumu}}

\def\rbrpsi{\ensuremath{r_{\mathrm{BR}}^{\psi}}\xspace}
\def\rbr{\ensuremath{r_{\mathrm{BR}}}\xspace}

\newif\ifEnableSectionTOCLinks
\EnableSectionTOCLinksfalse 

\input{LHCb/preamble}
\usepackage{longtable} 
\usepackage{makecell}
\usepackage{subcaption}

\begin{document}

\renewcommand{\thefootnote}{\fnsymbol{footnote}}
\setcounter{footnote}{1}

\input{title-LHCb-PAPER}


\renewcommand{\thefootnote}{\arabic{footnote}}
\setcounter{footnote}{0}


\cleardoublepage


\pagestyle{plain} 
\setcounter{page}{1}
\pagenumbering{arabic}


\input{body}

\input{LHCb/acknowledgements}


\input{appendix}



\addcontentsline{toc}{section}{References}
\bibliographystyle{LHCb/LHCb}
\bibliography{main,LHCb/standard,LHCb/LHCb-PAPER,LHCb/LHCb-CONF,LHCb/LHCb-DP,LHCb/LHCb-TDR,LHCb/LHCb-PUB}

\newpage
\input{LHCb/Authorship_LHCb-PAPER-2026-036}

\end{document}

%% file: LHCb/preamble.tex
\usepackage[top=1in, bottom=1.25in, left=1in, right=1in]{geometry}

\usepackage{microtype}
\usepackage{lineno}  
\usepackage{xspace} 
\usepackage{caption} 

\usepackage{graphicx}  
\usepackage{color}
\usepackage{colortbl}
\graphicspath{{./figs/}} 

\usepackage{amsmath} 
\usepackage{amssymb}
\usepackage{amsfonts}
\usepackage{upgreek} 

\usepackage[normalem]{ulem} 

\newcommand*\patchAmsMathEnvironmentForLineno[1]{%
\expandafter\let\csname old#1\expandafter\endcsname\csname #1\endcsname
\expandafter\let\csname oldend#1\expandafter\endcsname\csname
end#1\endcsname
 \renewenvironment{#1}%
   {\linenomath\csname old#1\endcsname}%
   {\csname oldend#1\endcsname\endlinenomath}%
}
\newcommand*\patchBothAmsMathEnvironmentsForLineno[1]{%
  \patchAmsMathEnvironmentForLineno{#1}%
  \patchAmsMathEnvironmentForLineno{#1*}%
}
\AtBeginDocument{%
\patchBothAmsMathEnvironmentsForLineno{equation}%
\patchBothAmsMathEnvironmentsForLineno{align}%
\patchBothAmsMathEnvironmentsForLineno{flalign}%
\patchBothAmsMathEnvironmentsForLineno{alignat}%
\patchBothAmsMathEnvironmentsForLineno{gather}%
\patchBothAmsMathEnvironmentsForLineno{multline}%
\patchBothAmsMathEnvironmentsForLineno{eqnarray}%
}

\usepackage[pdftex,
            pdfauthor={\paperauthors},
            pdftitle={\paperasciititle},
            pdfkeywords={\paperkeywords}]{hyperref}
\usepackage{hyperxmp}
\hypersetup{
    pdfcopyright={Copyright (C) \papercopyright},
    pdflicenseurl={\paperlicenceurl}
}

\usepackage[colorinlistoftodos,textsize=scriptsize]{todonotes}

\usepackage[bottom,flushmargin,hang,multiple]{footmisc}

\usepackage[all]{hypcap} 

\input{LHCb/lhcb-symbols-def} 

\hypersetup{
  colorlinks   = true, 
  urlcolor     = blue, 
  linkcolor    = blue, 
  citecolor    = red   
}

\ifEnableSectionTOCLinks
    \usepackage[explicit]{titlesec} 
    
    \let\oldcontentsline\contentsline
    \renewcommand

    \titleformat{\section}{\normalfont\Large\bf}{\hyperlink{tocsection.\thesection}{{\thesection} \parbox[t]{\dimexpr\textwidth-1pc}{#1}}}{1pc}{}

    \titleformat{\subsection}{\normalfont\bf}{\hyperlink{tocsubsection.\thesubsection}{{\thesubsection} \parbox[t]{\dimexpr\textwidth-1pc}{#1}}}{1pc}{}

    \titleformat{name=\section,numberless}[display]{}{}{0pt}{\normalfont\Huge\bfseries #1}
\fi

\usepackage{cite} 
\usepackage{LHCb/mciteplus}
\makeatletter
\g@addto@macro\bfseries{\boldmath}
\makeatother

%% file: LHCb/lhcb-symbols-def.tex
\usepackage{xspace} 
\usepackage{upgreek}

\def\lhcb   {\mbox{LHCb}\xspace}

\def\MagUp {\mbox{\em Mag\kern -0.05em Up}\xspace}

\ifdefined\ifuprightparticles
\else
\newboolean{uprightparticles}
\setboolean{uprightparticles}{false} 
\fi

\ifthenelse{\boolean{uprightparticles}}%
{

 \def\Pmu         {\ensuremath{\upmu}\xspace}

 \def\Ppi         {\ensuremath{\uppi}\xspace}

 \def\Ppsi        {\ensuremath{\uppsi}\xspace}

 \def\PDelta      {\ensuremath{\Delta}\xspace}                 
 \def\PXi         {\ensuremath{\Xi}\xspace}                 
 \def\PLambda     {\ensuremath{\Lambda}\xspace}                 
 \def\PSigma      {\ensuremath{\Sigma}\xspace}                 
 \def\POmega      {\ensuremath{\Omega}\xspace}                 
 \def\PUpsilon    {\ensuremath{\Upsilon}\xspace}
 \let\oldPi\Pi
 \def\PPi         {\ensuremath{\oldPi}\xspace}

 \def\PB      {\ensuremath{\mathrm{B}}\xspace}                 
 \def\PD      {\ensuremath{\mathrm{D}}\xspace}                 
 \def\PJ      {\ensuremath{\mathrm{J}}\xspace}                 
 \def\PK      {\ensuremath{\mathrm{K}}\xspace}                 
 \def\Pb      {\ensuremath{\mathrm{b}}\xspace}                 
 \def\Pc      {\ensuremath{\mathrm{c}}\xspace}                 
                  
 \def\Pe      {\ensuremath{\mathrm{e}}\xspace}                 
 \def\Ps      {\ensuremath{\mathrm{s}}\xspace}

 \def\thebaroffset{0.0em}
}
{

 \def\Pmu         {\ensuremath{\mu}\xspace}

 \def\Ppi         {\ensuremath{\pi}\xspace}

 \def\Ppsi        {\ensuremath{\psi}\xspace}                 
                  
 \mathchardef\PDelta="7101
 \mathchardef\PXi="7104
 \mathchardef\PLambda="7103
 \mathchardef\PSigma="7106
 \mathchardef\POmega="710A
 \mathchardef\PUpsilon="7107
 \mathchardef\PPi="7105
 \def\PB      {\ensuremath{B}\xspace}                 
 \def\PD      {\ensuremath{D}\xspace}                 
 \def\PJ      {\ensuremath{J}\xspace}                 
 \def\PK      {\ensuremath{K}\xspace}                 
 \def\Pb      {\ensuremath{b}\xspace}                 
 \def\Pc      {\ensuremath{c}\xspace}                 
                  
 \def\Pe      {\ensuremath{e}\xspace}                 
 \def\Ps      {\ensuremath{s}\xspace}

 \def\thebaroffset{0.18em}
}
\newcommand{\offsetoverline}[2][\thebaroffset]{\kern #1\overline{\kern -#1 #2}}%

\makeatletter
\ifcase \@ptsize \relax
  \newcommand{\miniscule}{\@setfontsize\miniscule{4}{5}}
\or
  \newcommand{\miniscule}{\@setfontsize\miniscule{5}{6}}
\or
  \newcommand{\miniscule}{\@setfontsize\miniscule{5}{6}}
\fi
\makeatother

\DeclareRobustCommand{\optbar}[1]{\shortstack{{\miniscule (\rule[.5ex]{1.25em}{.18mm})}
  \\ [-.7ex] $#1$}}

\def\epem       {{\ensuremath{\Pe^+\Pe^-}}\xspace}

\def\mumu       {{\ensuremath{\Pmu^+\Pmu^-}}\xspace}

\def\squark    {{\ensuremath{\Ps}}\xspace}

\def\cquark    {{\ensuremath{\Pc}}\xspace}
\def\cquarkbar {{\ensuremath{\overline \cquark}}\xspace}

\def\bquark    {{\ensuremath{\Pb}}\xspace}

\def\pion   {{\ensuremath{\Ppi}}\xspace}

\def\pim    {{\ensuremath{\pion^-}}\xspace}

\def\kaon    {{\ensuremath{\PK}}\xspace}

\def\KorKbar {\kern \thebaroffset\optbar{\kern -\thebaroffset \PK}{}\xspace}

\def\Kp      {{\ensuremath{\kaon^+}}\xspace}

\def\D       {{\ensuremath{\PD}}\xspace}

\def\DorDbar {\kern \thebaroffset\optbar{\kern -\thebaroffset \PD}\xspace}

\def\Dp      {{\ensuremath{\D^+}}\xspace}
\def\Dm      {{\ensuremath{\D^-}}\xspace}

\def\DpDm    {\ensuremath{\Dp {\kern -0.16em \Dm}}\xspace}

\def\B       {{\ensuremath{\PB}}\xspace}

\def\BorBbar {\kern \thebaroffset\optbar{\kern -\thebaroffset \PB}\xspace}

\def\Bd      {{\ensuremath{\B^0}}\xspace}

\def\BdorBdbar {\kern \thebaroffset\optbar{\kern -\thebaroffset \Bd}\xspace}
\def\Bu      {{\ensuremath{\B^+}}\xspace}

\def\Bp      {{\ensuremath{\Bu}}\xspace}

\def\Bs      {{\ensuremath{\B^0_\squark}}\xspace}

\def\BsorBsbar {\kern \thebaroffset\optbar{\kern -\thebaroffset \Bs}\xspace}

\def\jpsi     {{\ensuremath{{\PJ\mskip -3mu/\mskip -2mu\Ppsi}}}\xspace}
\def\psitwos  {{\ensuremath{\Ppsi{(2S)}}}\xspace}

\def\Y#1S{\ensuremath{\PUpsilon{(#1S)}}\xspace}

\def\Lz          {{\ensuremath{\PLambda}}\xspace}

\def\LorLbar     {\kern \thebaroffset\optbar{\kern -\thebaroffset \PLambda}\xspace}
\def\Lambdares   {{\ensuremath{\PLambda}}\xspace}

\def\Xires       {{\ensuremath{\PXi}}\xspace}
\def\Xiz         {{\ensuremath{\Xires^0}}\xspace}
\def\Xim         {{\ensuremath{\Xires^-}}\xspace}
\def\Xiresbar       {{\ensuremath{\offsetoverline{\Xires}}}\xspace}

\def\Omegares    {{\ensuremath{\POmega}}\xspace}

\def\Lb           {{\ensuremath{\Lz^0_\bquark}}\xspace}

\def\Xib          {{\ensuremath{\Xires_\bquark}}\xspace}
\def\Xibz         {{\ensuremath{\Xires^0_\bquark}}\xspace}
\def\Xibm         {{\ensuremath{\Xires^-_\bquark}}\xspace}

\def\Xibbarp      {{\ensuremath{\Xiresbar{}_\bquark^+}}\xspace}
\def\Omegab       {{\ensuremath{\Omegares^-_\bquark}}\xspace}

\def\BF         {{\ensuremath{\mathcal{B}}}\xspace}
\def\BR         {\BF}

\newcommand{\decay}[2]{\ensuremath{\mathinner{#1\!\to #2}}\xspace}

\def\to                 {\ensuremath{\rightarrow}\xspace}

\def\qsq       {{\ensuremath{q^2}}\xspace}

\def\bsll     {\decay{\bquark}{\squark \ell^+ \ell^-}}

\def\AT#1     {\ensuremath{A_{\mathrm{T}}^{#1}}\xspace}           

\def\C#1      {\ensuremath{\mathcal{C}_{#1}}\xspace}                       
\def\Cp#1     {\ensuremath{\mathcal{C}_{#1}^{'}}\xspace}                    
\def\Ceff#1   {\ensuremath{\mathcal{C}_{#1}^{\mathrm{(eff)}}}\xspace}        
\def\Cpeff#1  {\ensuremath{\mathcal{C}_{#1}^{'\mathrm{(eff)}}}\xspace}       
\def\Ope#1    {\ensuremath{\mathcal{O}_{#1}}\xspace}                       
\def\Opep#1   {\ensuremath{\mathcal{O}_{#1}^{'}}\xspace}                    

\newcommand{\nospaceunit}[1]{\ensuremath{\text{#1}}}       
\newcommand{\aunit}[1]{\ensuremath{\text{\,#1}}}       

\newcommand{\tev}{\aunit{Te\kern -0.1em V}\xspace}
\newcommand{\gev}{\aunit{Ge\kern -0.1em V}\xspace}
\newcommand{\mev}{\aunit{Me\kern -0.1em V}\xspace}
\newcommand{\kev}{\aunit{ke\kern -0.1em V}\xspace}
\newcommand{\ev}{\aunit{e\kern -0.1em V}\xspace}
 
\newcommand{\mevc}{\ensuremath{\aunit{Me\kern -0.1em V\!/}c}\xspace}
\newcommand{\gevc}{\ensuremath{\aunit{Ge\kern -0.1em V\!/}c}\xspace}
\newcommand{\mevcc}{\ensuremath{\aunit{Me\kern -0.1em V\!/}c^2}\xspace}
\newcommand{\gevcc}{\ensuremath{\aunit{Ge\kern -0.1em V\!/}c^2}\xspace}
\newcommand{\gevgevcccc}{\ensuremath{\gev^2\!/c^4}\xspace} 

\def\mum  {\ensuremath{\,\upmu\nospaceunit{m}}\xspace}

\def\fb   {\ensuremath{\aunit{fb}}\xspace}
\def\invfb   {\ensuremath{\fb^{-1}}\xspace}

\newcommand{\stat}{\aunit{(stat)}\xspace}
\newcommand{\syst}{\aunit{(syst)}\xspace}

\newcommand{\chisq}{\ensuremath{\chi^2}\xspace}

\newcommand{\chisqip}{\ensuremath{\chi^2_{\text{IP}}}\xspace}

\def\gsim{{~\raise.15em\hbox{$>$}\kern-.85em
          \lower.35em\hbox{$\sim$}~}\xspace}
\def\lsim{{~\raise.15em\hbox{$<$}\kern-.85em
          \lower.35em\hbox{$\sim$}~}\xspace}

\def\sPlot{\mbox{\em sPlot}\xspace}

\def\pt         {\ensuremath{p_{\mathrm{T}}}\xspace}

\def\ptot       {\ensuremath{p}\xspace}

\def\evtgen     {\mbox{\textsc{EvtGen}}\xspace}

\def\geant      {\mbox{\textsc{Geant4}}\xspace}

\def\photos     {\mbox{\textsc{Photos}}\xspace}

\def\pythia     {\mbox{\textsc{Pythia}}\xspace}

\def\tell1  {TELL1\xspace}
\def\ukl1   {UKL1\xspace}

\newcommand{\phz}{\phantom{0}}
\newcommand{\lhcborcid}[1]{\href{https://orcid.org/#1}{\hspace*{0.1em}\raisebox{-0.45ex}{\includegraphics[width=1em]{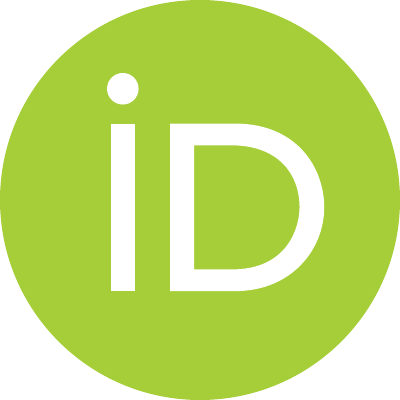}}}}


%% file: title-LHCb-PAPER.tex

\begin{titlepage}
\pagenumbering{roman}

\vspace*{-1.5cm}
\centerline{\large EUROPEAN ORGANIZATION FOR NUCLEAR RESEARCH (CERN)}
\vspace*{1.5cm}
\noindent
\begin{tabular*}{\linewidth}{lc@{\extracolsep{\fill}}r@{\extracolsep{0pt}}}
\ifthenelse{\boolean{pdflatex}}
{\vspace*{-1.5cm}\mbox{\!\!\!\includegraphics[width=.14\textwidth]{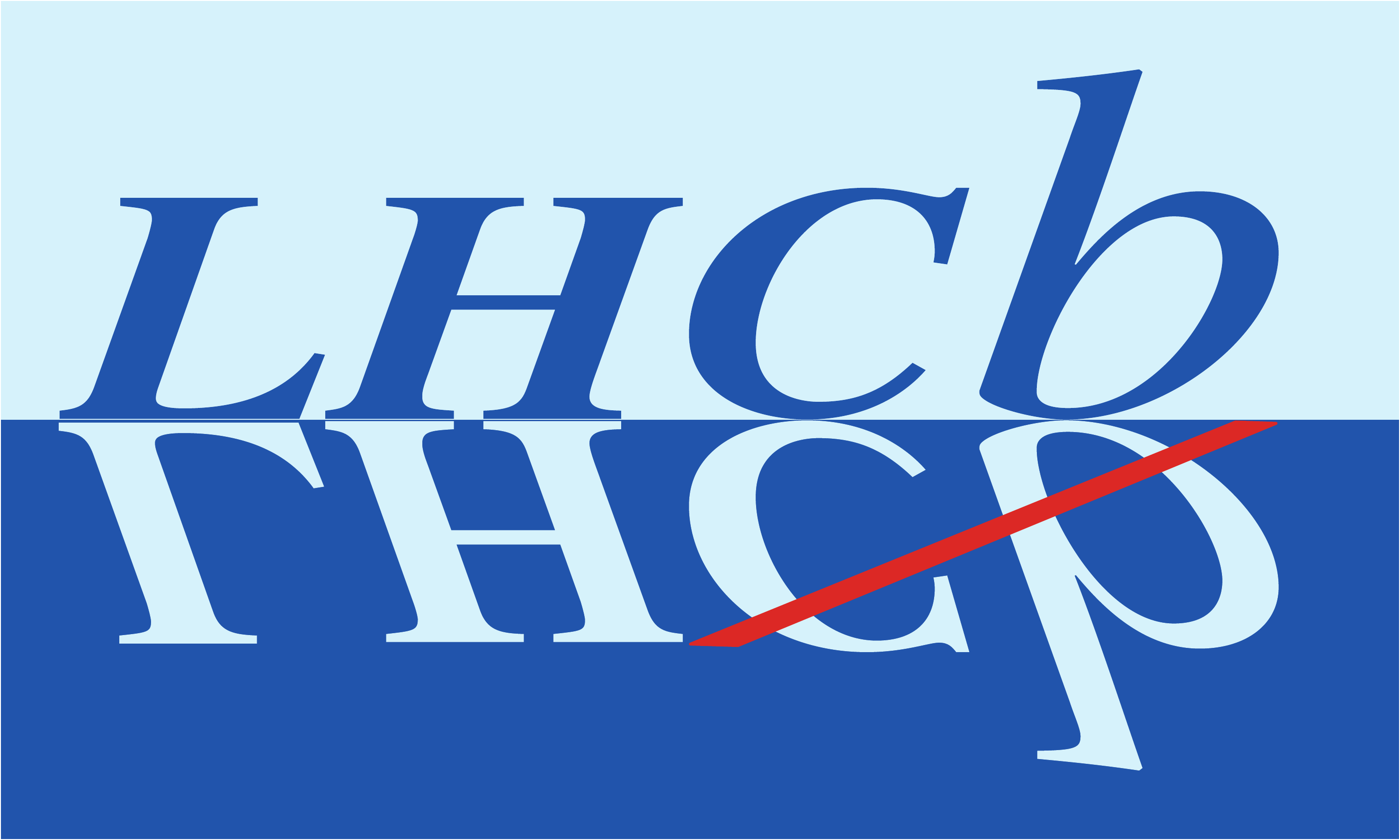}} & &}%
{\vspace*{-1.2cm}\mbox{\!\!\!\includegraphics[width=.12\textwidth]{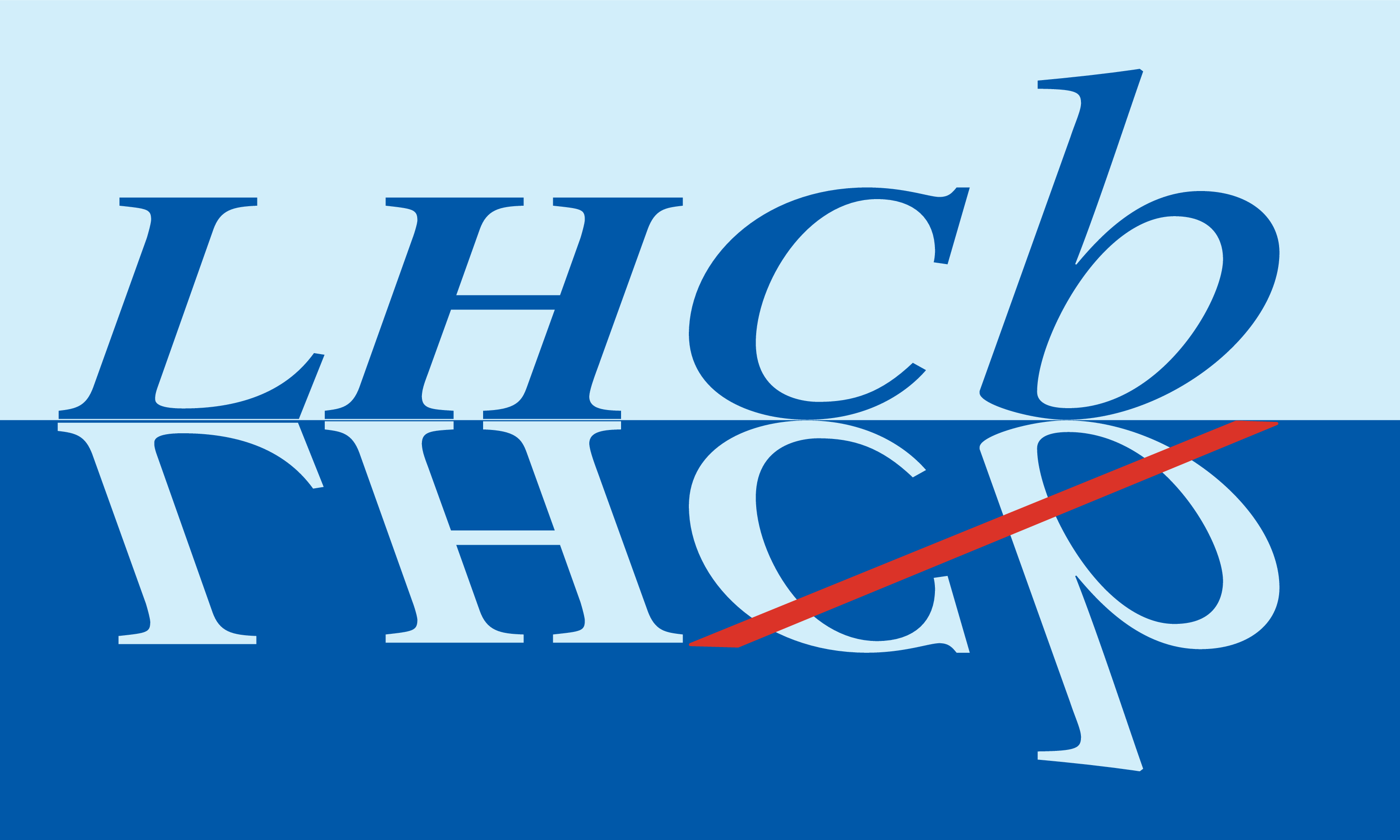}} & &}%
\\
 & & CERN-EP-2026-256 \\  
 & & LHCb-PAPER-2026-036 \\  
 & & 15 October 2026 \\ 
 & & \\
\end{tabular*}

\vspace*{4.0cm}

{\normalfont\bfseries\boldmath\huge
\begin{center}
  \papertitle 
\end{center}
}

\vspace*{2.0cm}

\begin{center}
\paperauthors\footnote{Authors are listed at the end of this paper.}
\end{center}

\vspace{\fill}

\begin{abstract}
  \noindent
The rare decay ${{\mathchar"7104}_b^-\to{\mathchar"7104}^-\mu^+\mu^-}$ is observed for the first time using proton-proton collision data collected by the LHCb experiment at centre-of-mass energies of $7$, $8$ and~$13\tev$, corresponding to integrated luminosities of $1$, $2$ and $6\invfb$, respectively.
Its branching fraction is reported in intervals of  the dimuon invariant mass squared.
The integrated value of $\BR({\mathchar"7104}_b^-\to{\mathchar"7104}^-\mu^+\mu^-)$ is measured to be ${ (118 \pm 24  \pm 7  \pm 38 ) \times 10^{-8}}$
, where the first uncertainty is  statistical, the second is  systematic and the third uncertainty arises from the normalisation branching fraction \({\cal B}({\mathchar"7104}_b^-\to{\mathchar"7104}^-J/\psi)\), whose uncertainty is dominated by SU(3)-breaking effects.
In addition, the ratio of the branching fractions of the ${{\mathchar"7104}_b^-\to{\mathchar"7104}^-\psi(2S)}$ and 
the ${{\mathchar"7104}_b^-\to{\mathchar"7104}^-J/\psi}$ decays is measured to be 
${ 0.491 \pm 0.041  \pm 0.019 }$.
  
\end{abstract}

\vspace*{2.0cm}

\begin{center}
  Submitted to JHEP 
\end{center}

\vspace{\fill}

{\footnotesize 
\centerline{\copyright~\papercopyright. \href{\paperlicenceurl}{\paperlicence}.}}
\vspace*{2mm}

\end{titlepage}


\newpage
\setcounter{page}{2}
\mbox{~}
%
%
%
%

%% file: body.tex
\section{Introduction}
\label{sec:Introduction}
Decays of $\bquark$ hadrons provide a sensitive probe for physics both within and beyond the Standard Model of particle physics (SM).
Flavour-changing neutral currents (FCNC), such as the \bsll transitions, where $\ell^\pm$ denotes a charged lepton, represent an especially compelling tool to test effects beyond the SM. In the SM, these processes can only be described through higher-order perturbative expansions, resulting in significant suppression of their rates. 
This rare occurrence makes \bsll processes sensitive to physics beyond the SM, with new physics effects potentially manifesting as deviations in decay rates or angular distributions from SM predictions.

Experimentally, hints of discrepancies with the SM expectations have been found in several measurements of differential decay rates~\cite{LHCb-PAPER-2021-014,LHCb-PAPER-2014-006} and angular observables~\cite{LHCb-PAPER-2025-041} of \bsll processes in \B-meson decays. 
The experimental and theoretical knowledge of baryonic \bsll transitions is considerably scarcer. Nevertheless, baryonic processes provide a complementary tool to study the deviations observed in the meson sector in greater depth, and offer additional  observables owing to the nonzero spin of baryons~\cite{Hiller:2021zth}. 

There are four baryons that contain one valence $b$ quark and two light quarks that decay through the weak interaction: the \Lb (valence quark content $udb$), \Xibm ($dsb$), \Xibz ($usb$) and \Omegab ($ssb$). Yet most measurements focused so far on decays of the \Lb baryon, as a consequence of its abundant production at the LHC~\cite{LHCb-PAPER-2018-050}, while the heavier baryons with strangeness have much lower production rates. For example, at the centre-of-mass energy $\sqrt{s}=13\tev$, the probability of a \bquark quark to hadronise into a \Xibm baryon is only about 8\% of that into a \Lb baryon~\cite{LHCb-PAPER-2018-047}. 

Several \bsll transitions have been observed for the \Lb baryon, namely, \LbToLmm~\cite{CDF:2011buy,LHCb-PAPER-2015-009}, \LbTopKmm~\cite{LHCb-PAPER-2016-059,LHCb-PAPER-2022-050} and \LbTopKee~\cite{LHCb-PAPER-2019-040}.
For the heavier \mbox{\bquark baryons}, no such processes have been searched for.
Despite the low production rate, the unique topology of the \XibToXimm decay chain, with the subsequent hyperon decays $\Xim \to \Lambdares\pi^-$ and $\Lambdares\to p\pi^-$, motivates the search for this decay mode.
Relatively low peaking background levels are expected since the decay topology cannot be mimicked by any meson decay due to the consecutive decays of the two long-lived hyperons, \Xim and \Lz. 

No model-independent measurements of the \Xibm production rate or branching fractions have been performed to date, with only a measurement of the product of these two quantities available~\cite{LHCb-PAPER-2018-047}. 
The $\Xibm \to \Xim$ transition is related to the $\Lb \to \Lz$ one via the approximate SU(3) flavour symmetry~\cite{Dery:2020lbc}, 
which predicts the \XibToXiJPsi decay width to be $3/2$ of that of \LbToLJPsi~\cite{Voloshin:2015xxa,Dery:2020lbc}. Similar considerations can be made for the ratio of decay widths of \XibToXimm and \LbToLmm decays~\cite{Wang:2021uzi}. A typical uncertainty of 30\% is often assumed on this ratio, accounting for the possible size of SU(3)-breaking corrections~\cite{LHCb-PAPER-2018-047}. A recent work that relies on lattice quantum chromodynamics (QCD) form factors provides the differential decay rate as a function of the dimuon invariant mass squared, \qsq~\cite{Farrell:2026swf}. Similarly to \LbToLmm, the \XibToXimm decays are expected to exhibit an enhancement at large values of dimuon invariant mass.  
Other theoretical predictions based on light-cone QCD as well as perturbative QCD predict integrated decay rates of about $10^{-6}$~\cite{Nayek:2020hna,Azizi:2011mw,Rui:2025ajk,Yadav:2024rax}. 
It is noteworthy that these theoretical expectations for the branching fraction of the \XibToXimm  decay are some of the largest among the \bsll processes observed to date. This may partially compensate for the low production rate of the \Xibm baryon, allowing the observation of this decay.

The \bsll transition receives contributions from resonant tree-level transitions $\bquark\to(\cquark\cquarkbar)s$, where 
$(\cquark\cquarkbar)$ denotes a charmonium state that decays into two leptons, most notably the \jpsi and \psitwos mesons. Being narrow states, these resonances appear as sharp peaks in the dimuon invariant-mass distribution.

This paper reports the first observation of the \XibToXimm decay and the first measurement of its branching fraction relative to the \XibToXiJPsi decay with \mbox{$\jpsi \to \mumu$}.\footnote{The inclusion of charge-conjugate processes is implied throughout.} This choice is made in order to increase the accuracy of the branching fraction measurement due to cancellation of the integrated luminosity, the $b$-quark production cross-section, the $\Xibm$ production fraction and other systematic uncertainties.
This choice additionally benefits from the possibility to fully align the selection thanks to the shared final-state particles and similar kinematics. The obtained ratio of decay rates is then reinterpreted to estimate the absolute branching fraction of the  \XibToXimm decay.

The ratio of the rates of \bquark-hadron decays into a strange hadron and either a \psitwos  or a \jpsi meson, is another interesting observable, since it enables the possibility of testing the role of nonfactorisable contributions to the $b\to (c\bar{c}) s$ transitions. In the \bquark-baryon sector, previous measurements of the ratio of \Lb decay rates into $\Lz\psitwos$ and $\Lz\jpsi$ final states, performed by the ATLAS and LHCb collaborations~\cite{Aad:2015msa,LHCb-PAPER-2017-041}, found a value of about 0.5. The experimental value agrees well with the perturbative QCD calculation~\cite{Rui:2022sdc}, but is in tension with respect to the nonrelativistic quark model~\cite{Cheng:1996cs} and the covariant quark model~\cite{Gutsche:2013oea,Gutsche:2015lea} that predict a value of about 0.8. 

A measurement of the ratio of  \XibToXiPsi and \XibToXiJPsi decay rates, with the $\psitwos$ and \jpsi mesons decaying to the dimuon final state, is also reported in this paper. 
The ratio has been measured by the CMS collaboration to be \mbox{$0.84^{\,+\,0.21}_{\,-\,0.19} \stat \pm 0.10 \,\syst \pm 0.02\, \textrm{(ext)}$}~\cite{CMS:2024rbi}, where the last uncertainty is due to the knowledge of the charmonia decay rates. The decays $\Xibz \to \Xiz\psitwos$ and $\Xibz \to \Xiz\jpsi$, related by the isospin symmetry to those studied in this paper, have been studied by LHCb recently~\cite{LHCb-PAPER-2026-030}, and the ratio of their rates is found to be $0.59\pm 0.19$.

The study of the $\Xim\mumu$ final state presented in this paper uses proton-proton ($pp$) collision data corresponding to a total integrated luminosity of 9~\invfb, comprising 3~\invfb collected at centre-of-mass energies of 7 and 8 \tev during 2011 and 2012 (Run 1) and 6~\invfb collected at 13 \tev from 2015 to 2018 (Run 2) with the LHCb detector.

\section{LHCb detector and simulation}
\label{sec:Detector}

The \lhcb detector~\cite{LHCb-DP-2008-001,LHCb-DP-2014-002} is a single-arm forward spectrometer covering the \mbox{pseudorapidity} range $2<\eta <5$, designed for the study of particles containing \bquark or \cquark quarks. The detector includes a high-precision tracking system consisting of a silicon-strip vertex detector (VELO) surrounding the $pp$
interaction region~\cite{LHCb-DP-2014-001}, a large-area silicon-strip detector located
upstream of a dipole magnet with a bending power of about $4{\mathrm{\,Tm}}$, and three stations of silicon-strip detectors and straw drift tubes~\cite{LHCb-DP-2013-003,LHCb-DP-2017-001} placed downstream of the magnet.
The tracking system provides a measurement of the momentum, \ptot, of charged particles with
a relative uncertainty that varies from 0.5\% at low momentum to 1.0\% at 200\gevc.
The minimum distance of a track to a primary $pp$ collision vertex (PV), the impact parameter (IP), 
is measured with a resolution of $(15+29/\pt)\mum$,
where \pt is the component of the momentum transverse to the beam, in\,\gevc.
Muons are identified by a system composed of alternating layers of iron and multiwire proportional chambers~\cite{LHCb-DP-2012-002}.

The online event selection is performed by a trigger~\cite{LHCb-DP-2012-004}, 
which consists of a hardware stage, followed by a software stage, which applies a full event reconstruction. At the hardware stage, the event is required to have a muon with a high \pt or a dimuon pair with a high product of \pt of the two muons. At the software stage, the dimuon pair must have a significant displacement from the PV.

Simulated samples of \XibToXimm, \XibToXiJPsi and \XibToXiPsi decays are used to model the detector effects and the selection requirements. 
In the simulation, $pp$ collisions are generated using \pythia~\cite{Sjostrand:2007gs} with a specific \lhcb configuration~\cite{LHCb-PROC-2010-056}. Decays of unstable particles are described by \evtgen~\cite{Lange:2001uf}, in which final-state radiation is generated using \photos~\cite{davidson2015photos}. The decays of the \Xibm baryon are simulated according to a phase-space distribution. The decays of the \Xim and \Lz hyperons are simulated with a dedicated \evtgen model, using the latest values of parity-violating asymmetry parameters~\cite{PDG2024}. The interaction of the generated particles with the detector, and its response, are implemented using the \geant toolkit~\cite{Allison:2006ve, *Agostinelli:2002hh} as described in Ref.~\cite{LHCb-PROC-2011-006}. 

The simulated properties of the dimuon in the \XibToXimm decays with the phase-space \Xibm decay model differ significantly from those observed in baryonic \bsll transitions. Therefore, using the expectation of SU(3) symmetry between the $\Xibm \to \Xim$ and $\Lb \to \Lz$ transitions, a sample of simulated \LbToLmm decays, generated using the form factors calculated by lattice QCD~\cite{Detmold:2016pkz}, is employed to correct the distributions of \qsq and two angular observables in the \XibToXimm decay. Those are the angle between the decay plane of the \Xim and the dimuon plane calculated in the \Xibm rest frame; and the angle between the positively charged muon direction in the dimuon rest frame and the dimuon momentum direction in the \Xibm rest frame. 
Further corrections are applied for known differences between data and simulation, as described in Sec.~\ref{sec:Efficiency}.

\section{Candidate selection}
\label{sec:Selection}

Owing to the long lifetime of the \Xim and \Lz hyperons, the $\Xim\mumu$ final state offers a cascade topology, which is distinct from most other processes occurring in the LHCb detector. This feature is exploited in the design of the selection requirements used in this analysis. They are based on the previous study of the decay \XibToXiJPsi at LHCb~\cite{LHCb-PAPER-2022-053}, but are modified to account for a worse signal-to-background ratio in the rare \XibToXimm decays. 
Reconstructed charged-hadron tracks are classified in two categories. The \textit{long-track} category refers to tracks that have reconstructed segments in both the VELO and the subsequent tracking stations. The \textit{downstream-track} category refers to those that are not reconstructed in the VELO, but only include information from the tracking detectors located just before and after the LHCb magnet. While most of the reconstructed charged particles produced in the $pp$ collisions form long tracks, most decay products of long-lived \Xim and \Lz hyperons are reconstructed as downstream tracks. Due to the presence of tracking information from the VELO detector, the trajectories, and hence the IP, of long tracks are measured with better precision than those of downstream tracks.

Signal \Xibm candidates are built by combining a \Xim hyperon candidate with two oppositely charged particles that are identified as muons by the muon system and are reconstructed as long tracks. The \Xim candidates are reconstructed in the $\Lz \pim$ decay mode. The \Lz hyperon candidate is built from a pair of oppositely charged particles, both forming either long or downstream tracks, that are assigned the proton and pion hypotheses, and originate from a common vertex displaced from the PV. 
The \Lz candidate is subsequently combined with a charged particle not originating from the PV, that is assigned a pion hypothesis, which can form either a long or downstream track (further denoted as $\pi_\PXi$). 
The \Xim hyperons are reconstructed in three track-type categories: three long tracks, two downstream tracks from the \Lz decay and one long track, or three downstream tracks. 
For the \XibToXiJPsi decay, the fractions of candidates in the three categories are approximately 10\%:45\%:45\%. 

All tracks used in this measurement are required to satisfy track-quality requirements by imposing a restriction on the probability that a track is reconstructed from a random set of hits. 
Both \Lz and \Xim candidates are required to have a good vertex-fit quality. The \Lz candidate must have an invariant mass within $8\mevcc$ around the known \Lz baryon mass~\cite{PDG2024}. The \Xim candidates are reconstructed constraining the mass of the \Lz candidate to its known value and required to have an invariant mass within $10\mevcc$ around the known value of the \Xim mass~\cite{PDG2024}.
The muon candidates are required to have $\pt > 800\mevc$, momentum $3 < p < 150\gevc$, and satisfy muon identification criteria. 
The \Xibm candidates are required to form a good vertex which is significantly displaced from the PV, and their momentum vectors are required to be consistent with originating from the PV. It is ensured that the two pion tracks are not duplicates by requiring that the angle between the two in the laboratory frame is greater than 0.5~mrad.

The \qsq interval [9.0, 9.9]\gevgevcccc is retained to select the \XibToXiJPsi candidates,\footnote{The dimuon invariant-mass resolution at the \jpsi peak in data is about 11\mevcc.} while the range [12.9, 14.0]\gevgevcccc is used to select the \XibToXiPsi candidates. 
The ranges [8.4, 9.0]\gevgevcccc and [9.9, 11.0]\gevgevcccc are not considered in this study, as they are contaminated by leakage from the \jpsi resonance due to final-state radiation and resolution effects.
The \qsq range  $[4m^2_\mu, 8.4 ] \cup [11.0, 12.5 ] \cup [14.0, (m_\Xibm - m_\Xim)^2]$, where $m_\mu$, $m_\Xibm$ and $m_\Xim$ are the known masses of the muon, \Xibm and \Xim, and the \qsq values are listed in \gevgevcccc, is used to search for the nonresonant \XibToXimm decay (Sec.~\ref{sec:Observation}).

Instead, for the measurement of the differential branching fraction of the \XibToXimm decay (Sec.~\ref{sec:BFrare}), the considered \qsq intervals are chosen to align with previous measurements in decays such as \LbToLmm~\cite{LHCb-PAPER-2015-009} and theoretical predictions.
A set of dimuon invariant-mass regions (denoted as I -- III) is defined in terms of \qsq and is summarised in Table~\ref{tab:qsq_windows}.

\begin{table}[b]
  \centering
  \caption{Dimuon invariant-mass squared regions used for the differential branching fraction measurement. }
  \label{tab:qsq_windows}
  \begin{tabular}{c|c}

    Region & \qsq range [\gevgevcccc]\\
\hline
I & $\qsq \in [\phz0.1, \phz8.0]$\rule{0pt}{0.9\normalbaselineskip}\\
\jpsi  & $\qsq \in [\phz9.0, \phz9.9]$  \\
II  & $\qsq \in [11.0, 12.5] $ \\
\psitwos  & $\qsq \in [12.9, 14.0]$  \\
III & $\qsq \in [15.0, 20.0] $\\

 \end{tabular}
\end{table}

In order to suppress the background from \LbToLmm decays combined with a random pion, the $\pi_\PXi$ candidate is required to have a significant displacement from the PV. 
The dominant partially reconstructed background component $\Xib \to \Xires(1530) \mumu$, where the $\Xires(1530)$ resonance decays to a $\Xim \pi$ final state and the dimuon is either nonresonant or originates from a charmonium resonance, is expected to be suppressed in any of the signal decays studied in this paper~\cite{Oliver:2010im}. This has recently been confirmed experimentally in the resonant \jpsi channel~\cite{LHCb-PAPER-2024-049}.
No significant partially reconstructed background is therefore expected close to the \Xibm mass peak.

Further suppression of the background made of random track combinations (combinatorial background) is achieved by using a Boosted Decision Tree (BDT) algorithm~\cite{Breiman,AdaBoost}, which exploits the gradient boosting technique implemented in the TMVA toolkit~\cite{Hocker:2007ht,*TMVA4}. The BDT classifier is constructed using kinematic and geometric properties, as well as the \chisqip of the final-state muons, the \Xim candidate and its decay products; along with the quality of the vertex fit of the \Xim candidate. Here, the \chisqip quantifies the change in the vertex-fit \chisq when the considered track is included in the PV fit. 
In addition, \pt, \chisqip, the quality of the vertex fit, and the angle between the direction of flight and the momentum of the \Xibm candidate are employed. The classifier is trained using the simulated nonresonant \XibToXimm decays as signal proxy, corrected for known differences between the data and simulation, as described below. The background proxy is formed by selecting the candidates in the sidebands of the data $\Xim\mumu$ invariant-mass distribution, in the intervals \mbox{(5000, 5600)} and \mbox{(5900, 7000)}\mevcc, in the \qsq intervals that exclude the charmonia resonances. 
To avoid potential biases, a cross-validation k-folding technique is used~\cite{10.1145/307400.307439}, with $k=10$. The requirement on the BDT output is optimised by using the figure of merit of Ref.~\cite{Punzi:2003bu}. This is done separately for the \XibToXimm and \XibToXiPsi decays, as the latter shows a significantly lower background contamination. For the same reason, the requirement is optimised independently for the \XibToXimm candidates with \qsq above the \psitwos resonance. For the study of the \XibToXiPsi decay, the same BDT output requirement is imposed on the normalisation \XibToXiJPsi candidates to allow for partial cancellation of the BDT efficiency. However, for the analysis of the \XibToXimm candidates, the \XibToXiJPsi candidates are assigned the same BDT requirement as the \XibToXimm candidates below the \psitwos resonance. This ensures that possible inaccuracies in the knowledge of the efficiency of the BDT requirement cancel in the ratio of the signal and normalisation decay modes. 

When analysing the $\Xim\mumu$ candidates that do not originate from an intermediate charmonium resonance, those within the invariant-mass window (5737, 5857)\mevcc were not examined until the entire analysis procedure was established. 

\section{Determination of the signal efficiencies}
\label{sec:Efficiency}
The efficiencies for signal \Xibm candidates 
are estimated using the corrected simulation samples of the respective decay modes. These efficiencies account for the LHCb detector acceptance, the detection, reconstruction, trigger and muon identification efficiency as well as the final kinematic,  geometric and multivariate selection. 
In order to improve the agreement between data and simulation in the three decay channels, the simulated \pt distribution of the \Xibm baryons is corrected using the \XibToXiJPsi data after the background contribution is subtracted using the \sPlot method~\cite{Pivk:2004ty}.
The momentum asymmetries of the decay products of the \Xim and \Lambdares hyperons, calculated in the laboratory frame, are corrected in order to adjust the angular distributions in the simulation of the \Xibm decays, which are modelled according to the phase-space distribution. The derived momentum asymmetry correction from the \jpsi mode is only applied to the \XibToXiJPsi and \XibToXiPsi simulated samples.

Then all three decay channels are also corrected for the hardware trigger response based on a tag-and-probe approach using $\Bp \to \jpsi \Kp$ decays~\cite{LHCb-DP-2012-004}. Similarly, the muon identification response is corrected with dedicated tag-and-probe $\jpsi \to \mumu$ samples~\cite{LHCb-DP-2018-001}. Lastly, in the simulated samples that correspond to Run 1 data-taking conditions, the vertexing efficiency of downstream-track combinations is weighted to match that measured in data. 
After the full correction, for a large number of variables that describe the decay kinematics and topology, proportion of track-type categories in \Xim decay, as well as the BDT response, a good agreement between the \XibToXiJPsi simulation and background-subtracted \XibToXiJPsi data is observed.

The presence of long-lived hyperons that often decay outside of the LHCb acceptance reduces the efficiency by about an order of magnitude. 
Furthermore, the efficiency is found to exhibit a strong dependence on \qsq, as observed in previous LHCb measurements that involve hyperons~\cite{LHCb-PAPER-2015-009}. This is explained by the anticorrelation between \qsq and the hyperon boost, affecting the probability of the hyperon decay to happen in the acceptance of the LHCb tracking system. Moreover, the trigger efficiency grows with the muon \pt, which is correlated to \qsq.
As a result, the ratio of efficiencies between the \XibToXiPsi and \XibToXiJPsi decays is about 1.32 in both Run~1 and Run~2; uncertainties on these ratios are discussed in the following Section. 

\begin{figure}[!htb]
        \centering
        \includegraphics[height=2.6in]{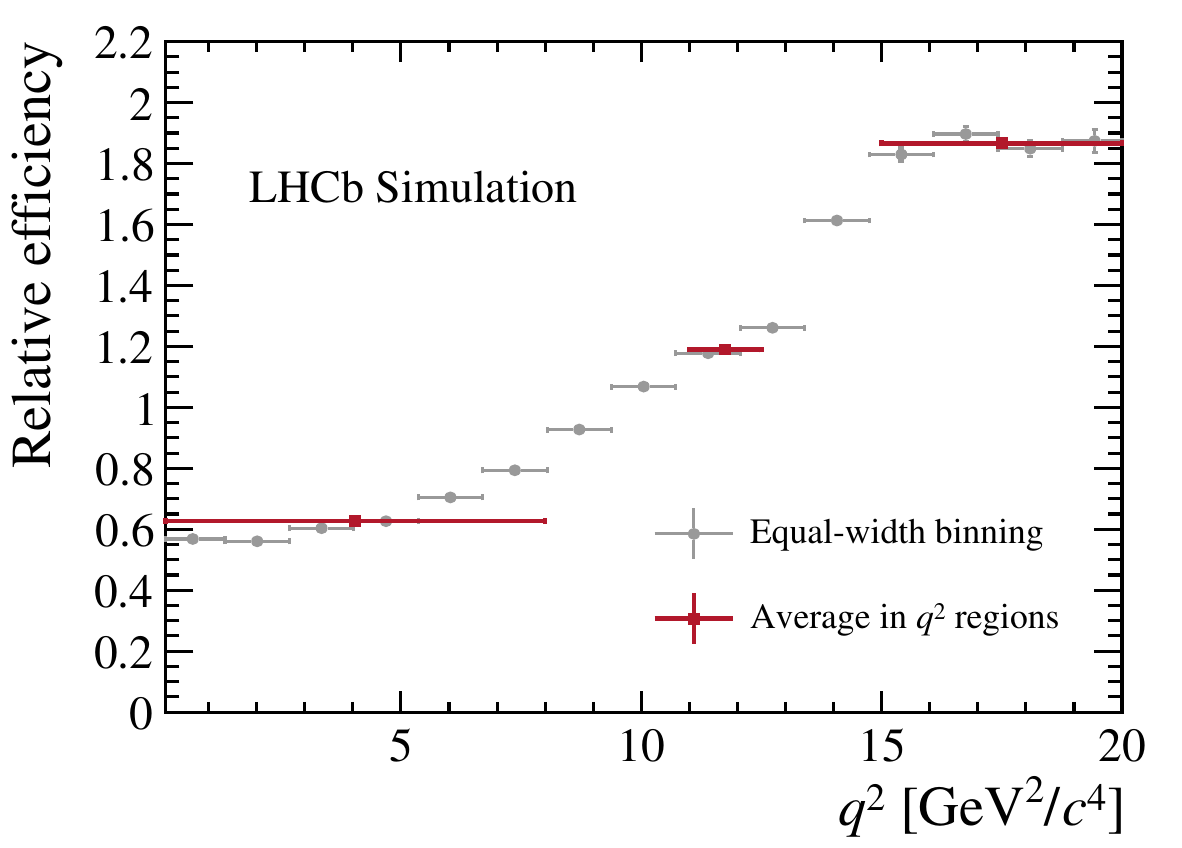}

    \caption{Ratio of the total \XibToXimm efficiency relative to that of the \XibToXiJPsi decay, as a function of \qsq. The red markers indicate the average value for each \qsq regions (I -- III) as described in Table~\ref{tab:qsq_windows}. The uncertainties only reflect the limited size of the simulation sample. The Run 1 and Run 2 data-taking periods are merged.}
    \label{fig:tot_eff_trends}
\end{figure}

The ratio of efficiencies of the \XibToXimm and \XibToXiJPsi decays is studied as a function of \qsq of the former. For the measurement of the differential \XibToXimm decay rate, only the average efficiency in each given \qsq interval is considered, as shown in Fig.~\ref{fig:tot_eff_trends}.
This ratio notably takes into account the \qsq dependence of the reconstruction and selection efficiency, as explained above. 
Additional sources of dependence arise from the trigger efficiency being lower for low-momentum muons, as well as the looser BDT selection requirement employed at high \qsq values.

\section{Measurement of the ratio of \XibToXiPsi and \XibToXiJPsi decay rates}
\label{sec:BFpsi}

The ratio 
\begin{equation}
\begin{split}
 \rbrpsi \equiv &\frac{\BR(\XibToXiPsi)}{\BR(\XibToXiJPsi)} = 
 \\
 &\frac{N(\XibToXiPsi)}{N(\XibToXiJPsi)} \times \frac{\varepsilon(\XibToXiJPsi)}{\varepsilon(\XibToXiPsi)} \times \frac{\BR(\jpsi\to\mumu)}{\BR(\psitwos\to\mumu)}
 \label{eq:r_psi_br}
\end{split}
\end{equation}
is measured, where $N$ indicates the measured yields and $\varepsilon$ the respective total efficiencies obtained from fully corrected simulation. The branching fractions, \BR, of the two charmonia decay modes are taken from Ref.~\cite{PDG2024}.
The collected data are considered separately for the Run 1 and Run 2 data-taking periods, which have slightly different invariant-mass resolutions. 
The value of \rbrpsi is extracted from an extended unbinned maximum-likelihood fit to the $\Xim\jpsi$ and $\Xim\psitwos$ invariant masses. In this fit, the $\Xim\jpsi$ and $\Xim\psitwos$ invariant masses are calculated by constraining the \Xim and \Lz invariant masses, as well as the charmonia resonances, to their known values~\cite{PDG2024}, and constraining the \Xibm candidate to originate from the PV.
The fit is performed simultaneously to the two data-taking periods and the two decay channels. It is implemented in the \textsc{RooFit} toolkit~\cite{Verkerke:2003ir} within the \textsc{ROOT} framework~\cite{Brun:1997pa}. 
Each signal parametrisation is described by a linear combination of two double-sided Crystal Ball functions~\cite{Skwarnicki:1986xj} that share a common mean. 
The value of mass resolution is taken from simulation, scaled by a factor of 1.1 determined using the \XibToXiJPsi signal in data and simulation.
Additionally, the tail parameters of the two Crystal Ball distributions and the relative fraction of these two components have been fixed to the values obtained from simulation.
In the simultaneous fit, the \rbrpsi and the \Xibm mass are shared parameters between the considered datasets.
The free parameters  in the fit to data are  \rbrpsi, the \Xibm mass, and the yields of \XibToXiJPsi decays in each data-taking period. Furthermore, the background is described in each dataset by an exponential function, with a freely varying normalisation and slope. 
The yields of the \XibToXiPsi decays in each data-taking period are expressed via \rbrpsi, the corresponding yields of the \XibToXiJPsi decays and the ratios of efficiencies of the two decay modes in the given data-taking period.

\begin{figure}[t]
    \centering
    \begin{subfigure}[t]{0.5\textwidth}
        \includegraphics[height=2.8in]{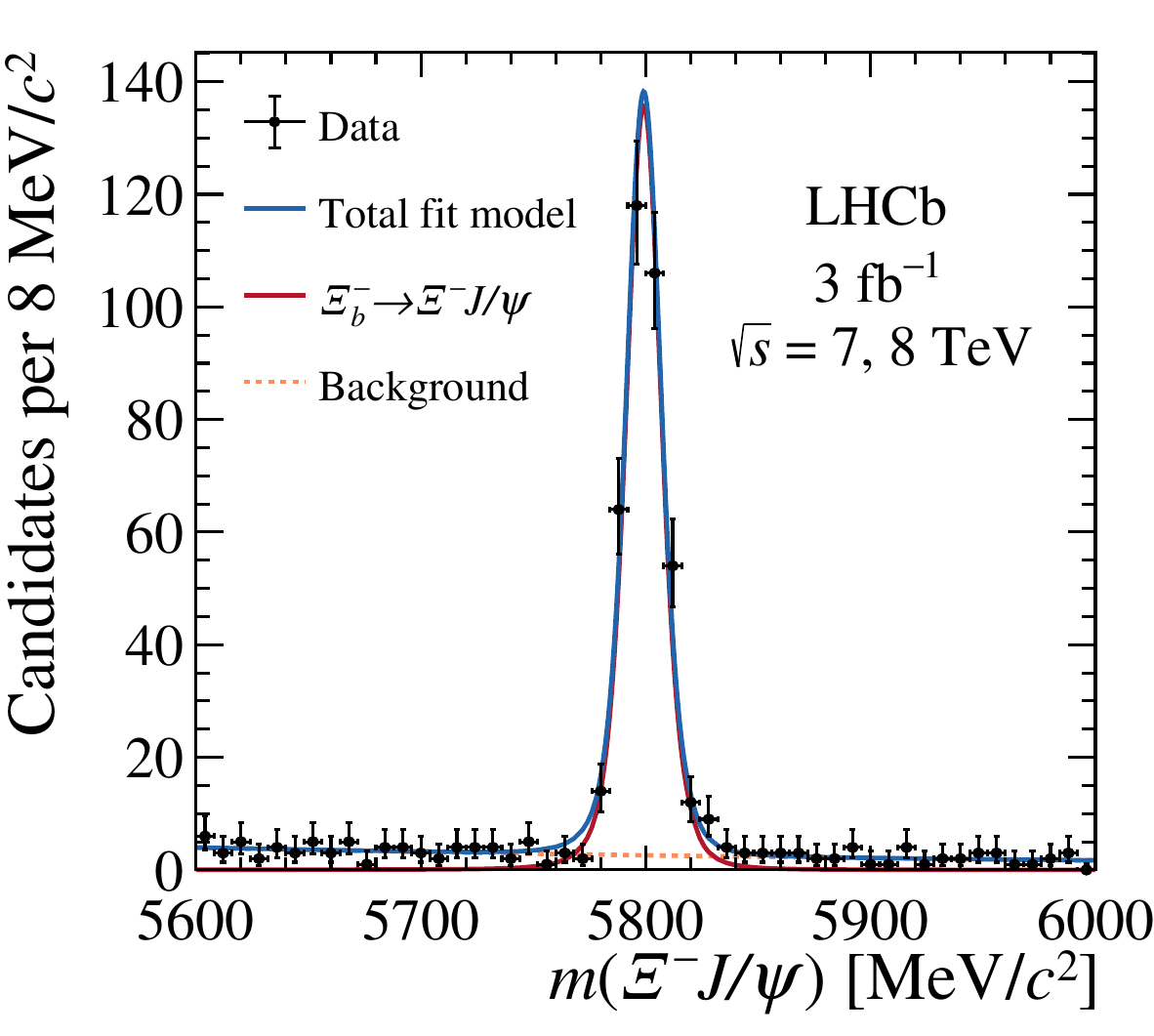}
    \end{subfigure}%
    \begin{subfigure}[t]{0.5\textwidth}
        \includegraphics[height=2.8in]{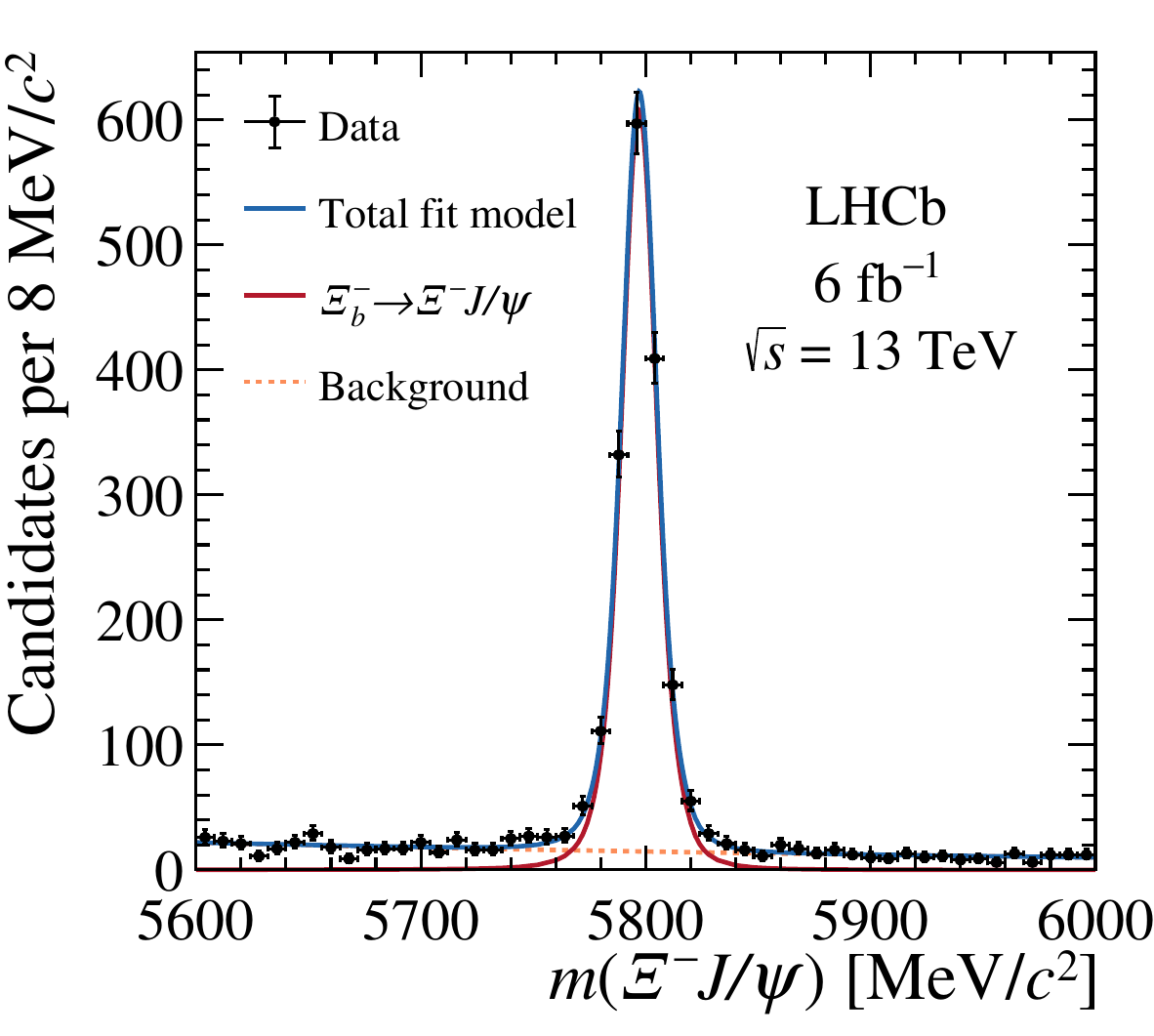}
    \end{subfigure}

    \begin{subfigure}[t]{0.5\textwidth}
        \includegraphics[height=2.8in]{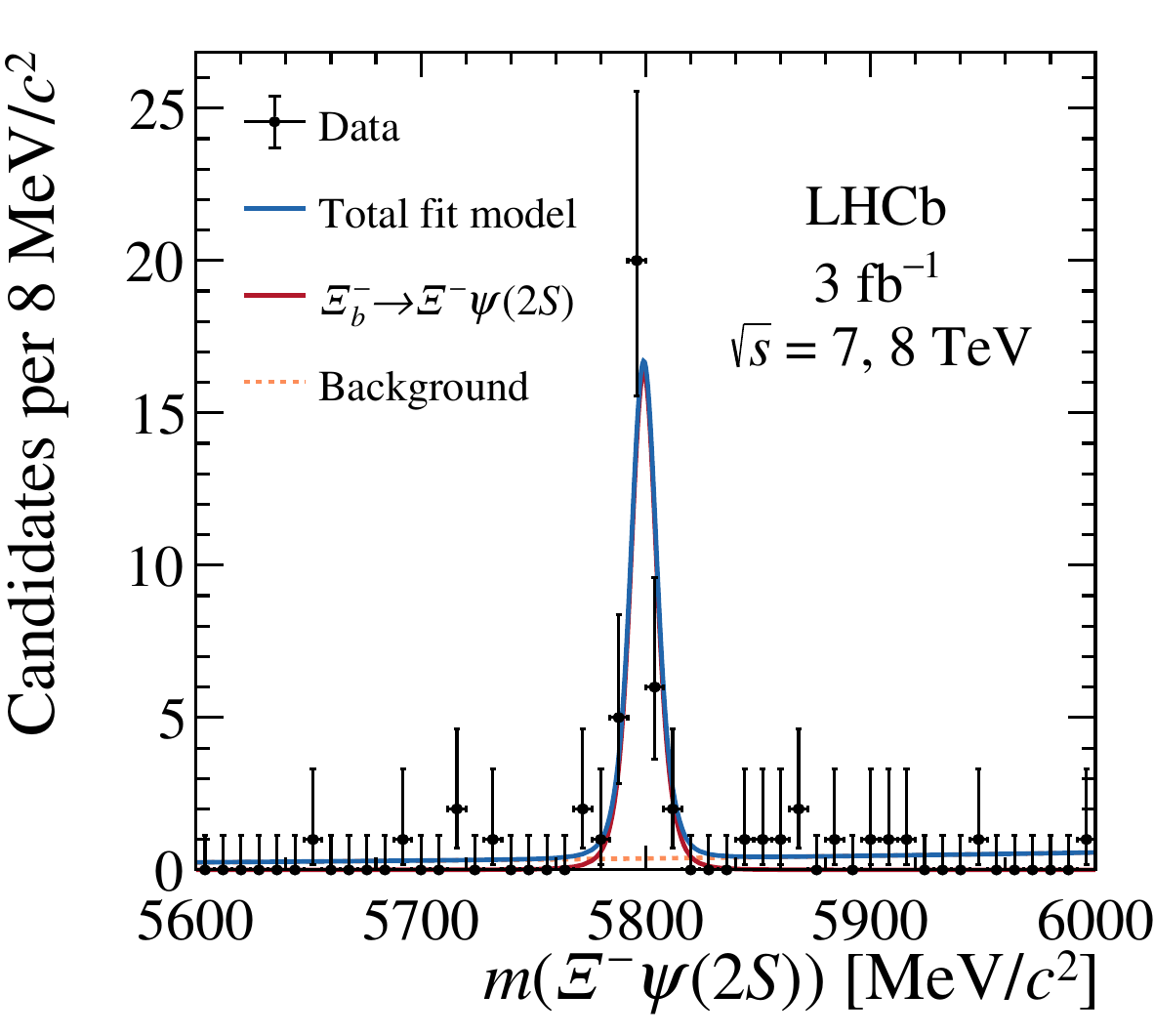}
    \end{subfigure}%
    \begin{subfigure}[t]{0.5\textwidth}
        \includegraphics[height=2.8in]{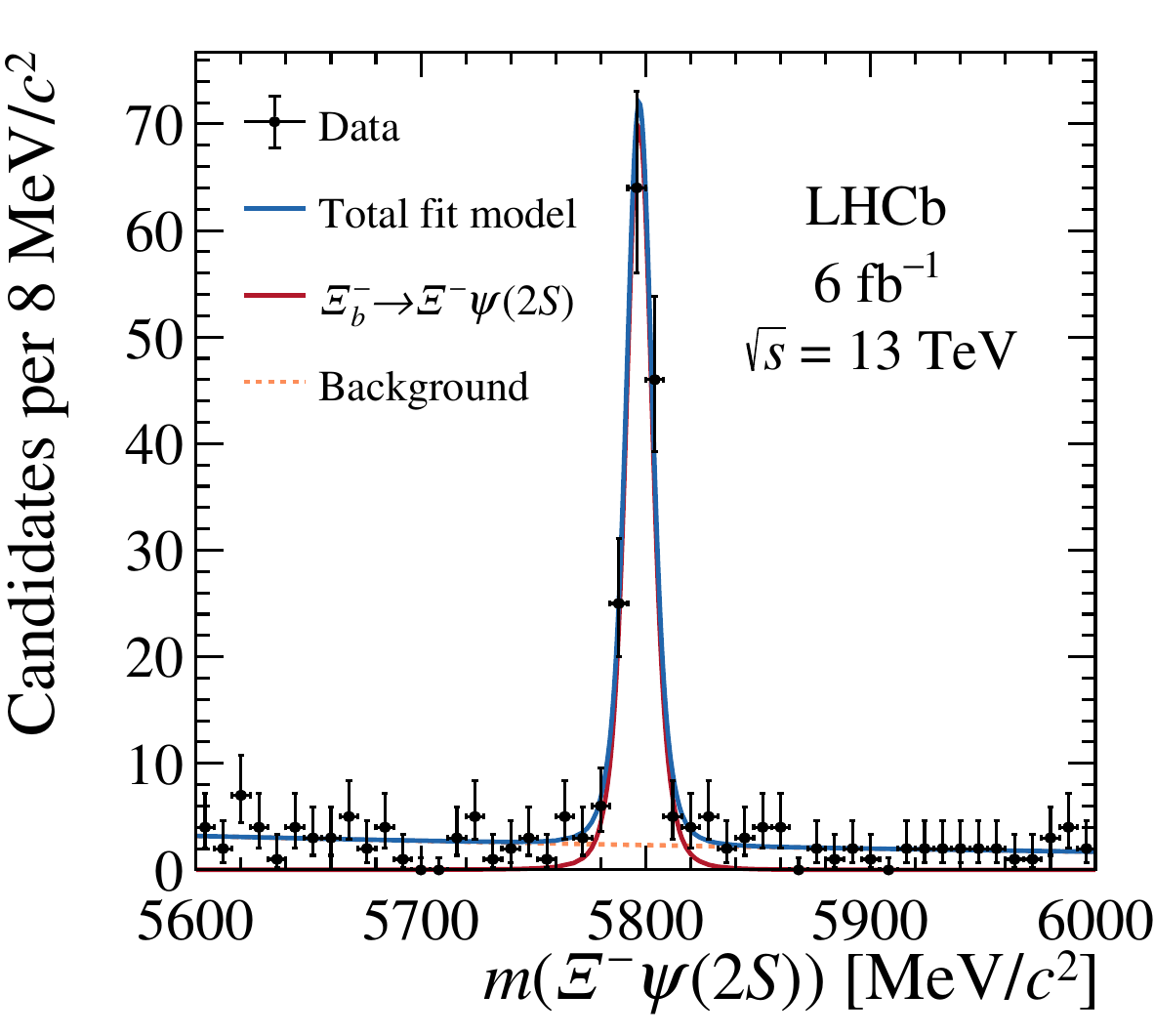}
    \end{subfigure}

    \caption{Invariant-mass distributions in the (top) \XibToXiJPsi datasets and (bottom) \XibToXiPsi datasets in (left) Run 1 and (right) Run 2. The results of the simultaneous fit are overlaid, with the (full red line) signal and (dotted line) background components also shown. }
    \label{fig:simfit_psi2s_wsyst}
\end{figure}

The resulting yields of the \XibToXiJPsi decay are $368\pm19$ and $1656\pm43$ in the Run~1 and~2 datasets, respectively. The value of \rbrpsi is found to be $0.491 \pm 0.041 \stat$. 
The selected datasets and the fit result are shown in Fig.~\ref{fig:simfit_psi2s_wsyst}. 
The yields of \XibToXiPsi decay are evaluated to be $33\pm 6$ and $141\pm 13$ in the Run 1 and 2 datasets, respectively.
To demonstrate robustness of the obtained result, the sample is split between different data-taking periods, between \Xibm and \Xibbarp decays and in the three track-type categories. Each individual result is found to be within 1.5 standard deviations from the default value. 

Possible sources of systematic uncertainty are studied separately for Run 1 and Run 2 data-taking periods and summarised in Table~\ref{tab:syst_rpsi}. Depending on the origin of each source of uncertainty, they are assumed to be either fully correlated or fully uncorrelated between the two data-taking periods. 
\begin{table}[t]
  \centering
  \caption{Relative systematic uncertainties on \rbrpsi. A value per data-taking period is 
  provided for the uncorrelated uncertainties, while a single value is given for the correlated ones. }
  \label{tab:syst_rpsi}
  \begin{tabular}{c|cc}

   Type & Run 1 [\%] &  Run 2 [\%]\\
\hline
  Size of simulated samples &  $0.6$ & $0.4$ \\
  Correction of simulation &  $1.2$ & $1.2$  \\
  Downstream tracking &  $5.2$ & $1.5$ \\
  Lifetime of \Xibm &  \multicolumn{2}{c}{$0.1$} \\
  Material interactions  &  \multicolumn{2}{c}{$1.1$} \\
  Fit model &  \multicolumn{2}{c}{$1.4$} \\
  External input ($\mathcal{B}$) &  \multicolumn{2}{c}{$2.8$}\\
       \hline 
   {Total uncorrelated} & $5.4$ & $2.0$ \\
   {Total correlated} & \multicolumn{2}{c}{$3.4$} \\
        \hline
   {Total} & 6.4 & 3.9 \\   \\ 
 \end{tabular}
\end{table}

The size of the simulated samples, used to estimate the signal efficiencies, leads to a $0.6\%$ ($0.4\%$) relative uncertainty on the efficiency ratio in Run 1 (Run 2). 
The systematic uncertainty associated with the simulation correction of kinematic distributions is assessed using a bootstrapping technique~\cite{efron:1979}. All the data candidates used for the determination of the correction weights are assigned a Poisson-distributed weight of unity mean. 
The correction procedure is repeated 100 times, resulting in 100 alternative efficiency-ratio values. These are then utilised to calculate the standard deviation of the efficiency ratio which is assigned as systematic uncertainty. 
Furthermore, alternative approaches are tested for the evaluation of the correction weights on the simulation. 
For instance, the kinematics of the \Xibm baryon is alternatively corrected in two dimensions, $\pt(\Xibm)$ and $\eta(\Xibm)$, rather than the default correction of $\pt(\Xibm)$ alone. The impact of applying the momentum-asymmetry correction is studied to assess the potential effect of mismodelling of the angular distribution.  For the correction of the hardware trigger response, an alternative calibration dataset is used; in addition, the separate correction of single-muon and dimuon triggers is compared to the simultaneous correction of the two. The impact of applying certain corrections derived from the Run 2 dataset to the Run 1 dataset is tested by extracting them from the Run 1 dataset; the observed difference in the efficiency ratio is negligible. 
A dedicated systematic uncertainty is also assigned for the correction procedure of the muon identification efficiency in simulation (0.6\%)~\cite{LHCb-PUB-2016-021}. Overall, the systematic uncertainty related to the correction of the simulation is found to be $1.2\%$ in both data-taking periods.

A further source of uncertainty originates from imprecise modeling of the downstream tracking and vertexing efficiency in simulation. This effect is expected to largely cancel in the ratio of the two decay modes. 
To exclude a momentum-dependent bias, the result is tested as a function of the \Xim momentum, and the observed variation is found to be within 1\% (5\%) for the Run 2 (Run 1) datasets. 
This is in agreement with the estimates from previous studies based on larger datasets of \LbToLJPsi decays~\cite{LHCb-PAPER-2025-035}.
Additional studies are performed by tightening the selection on the probability that a track is reconstructed from a random set of hits for downstream tracks, to exclude potential mismodelling in this variable. As a result, a systematic uncertainty of $1.5\%$ ($5.2\%$) is assigned for Run 2 (Run 1) dataset. 

A further uncertainty arises from the fact that the value of the \Xibm baryon lifetime used to generate the simulated samples differs, by less than one standard deviation, from the latest LHCb measurement~\cite{LHCb-PAPER-2024-010}. The corresponding systematic uncertainty is determined by weighting the decay-time distributions in the simulation with an exponential factor that reflects the change in the lifetime value. This results in a $0.1\%$ relative uncertainty on the efficiency ratio. 

Long-lived \Xim and \Lz hyperons can interact with the detector material prior to their decay. 
To estimate the potential bias originating from the modelling of the \qsq dependence of such material interactions in simulation, the probability to survive material interactions is studied as a function of \qsq using  \XibToXimm simulated decays, and is found to be consistent with being constant, with the value of $(95\pm 1)\%$. 
However, the knowledge of the interaction cross-sections for hyperons is rather poor. 
Using the hyperon material-interaction probability estimated with two different versions of the \geant simulation toolkit (\texttt{9.6.4} and \texttt{10.6.2}), which use different values for hadronic interaction cross-sections, amounts to a relative difference of $0.5\%$.  Summing these two effects in quadrature, the uncertainty of~$1.1\%$ is assigned to account for the impact of hyperon interaction with material. 

Several sources of systematic uncertainty related to the fit model are considered. The uncertainties related to the fit bias and the chosen signal and background models are studied using pseudoexperiments. In particular, the possible background from \LbToLJPsi (\LbToLPsi) decays combined with a random pion, that leads to the same final state as \XibToXiJPsi (\XibToXiPsi), is examined. According to simulation, this \Lb component has a broad shape in the $\Xi\mumu$ invariant mass that starts at the kinematic threshold $m(\Lb)+m(\pim)$ and has a long tail to higher invariant masses. 
When modelled using kernel-density estimates and included in the fit, the \LbToLJPsi (\LbToLPsi) components have yields consistent with zero. 
Since the rates of the corresponding \LbToLJPsi (\LbToLPsi) and \XibToXiJPsi (\XibToXiPsi) decays can be linked by the flavour SU(3) symmetry, while the ratio of the \LbToLPsi and \LbToLJPsi decay rates is known, one can estimate the range of expected \Lb background yields in each dataset. These estimates are then used to generate pseudoexperiments, which are fitted with the default fit model. The impact of the possible \Lb background on the measurement of \rbrpsi is found to be at sub-percent level. Further sources of uncertainty are estimated due to the modelling of the combinatorial background and the signal shape, by generating pseudoexperiments with alternative models and fitting them with the default one. 
The final relative systematic uncertainty from the fit model is $1.4\%$. 

Lastly, the uncertainties on the branching fractions of the dimuon decays of the charmonium are propagated to the final result, leading to a relative uncertainty of $2.8\%$. For the $\psitwos \to \mumu$ decay, lepton universality is assumed, and the dimuon decay rate is assumed to match the dielectron decay rate, which is known more precisely. 
Accounting for the fit result and the systematic uncertainties added in quadrature, the ratio is measured to be
\begin{equation*}
 \rbrpsi = 0.491 \pm 0.041 \stat \pm 0.019 \syst.
 \label{eq:result_rbrpsi}
\end{equation*}

The profile log-likelihood scan for the \rbrpsi observable is shown in App.~\ref{app:pll}.

\section{Observation of the \XibToXimm signal}
\label{sec:Observation}

\begin{figure}[t]
    \centering
    \begin{subfigure}[t]{0.5\textwidth}
        \centering
        \includegraphics[height=2.8in]{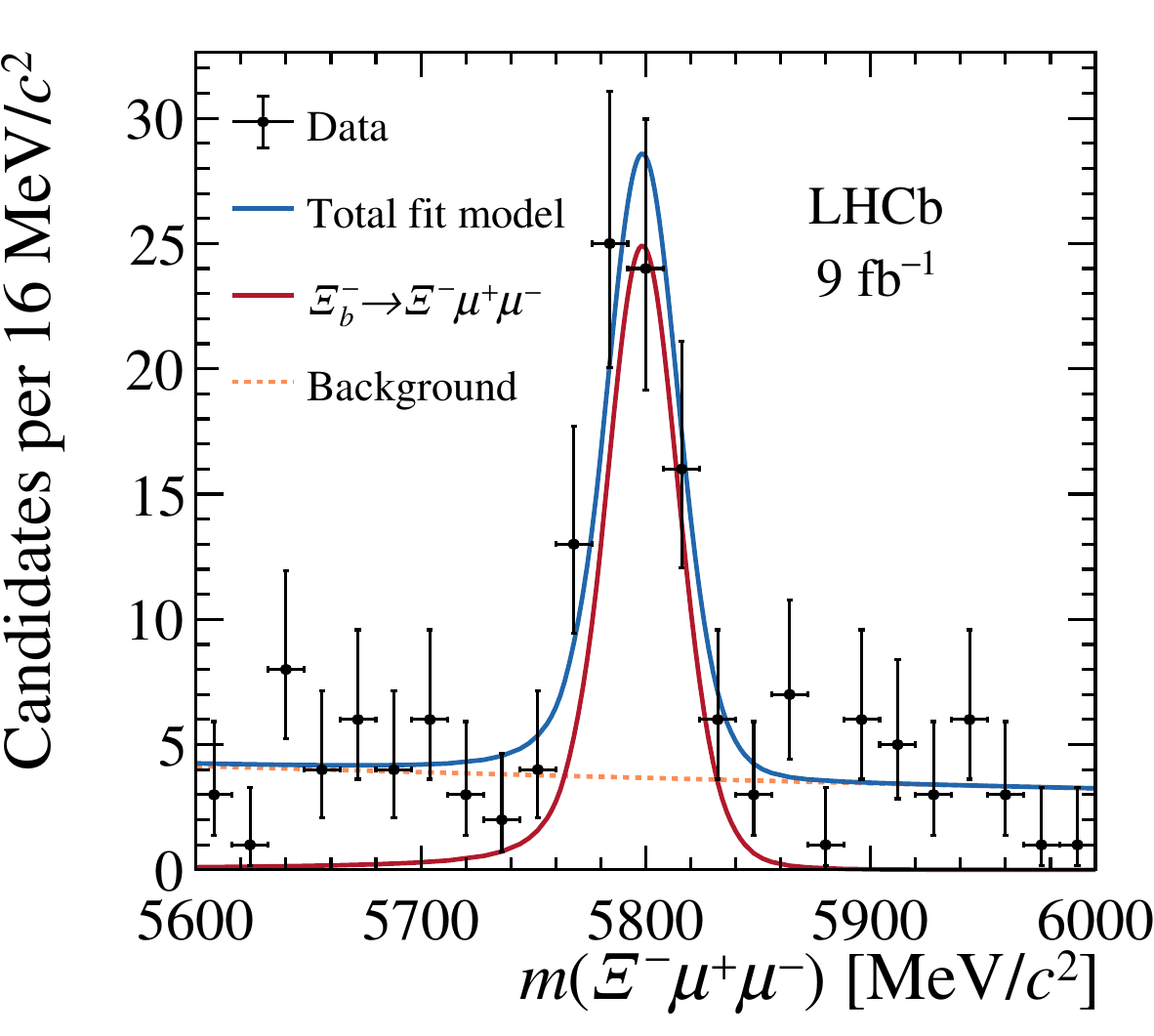}
    \end{subfigure}%
    \begin{subfigure}[t]{0.5\textwidth}
        \centering
        \includegraphics[height=2.8in]{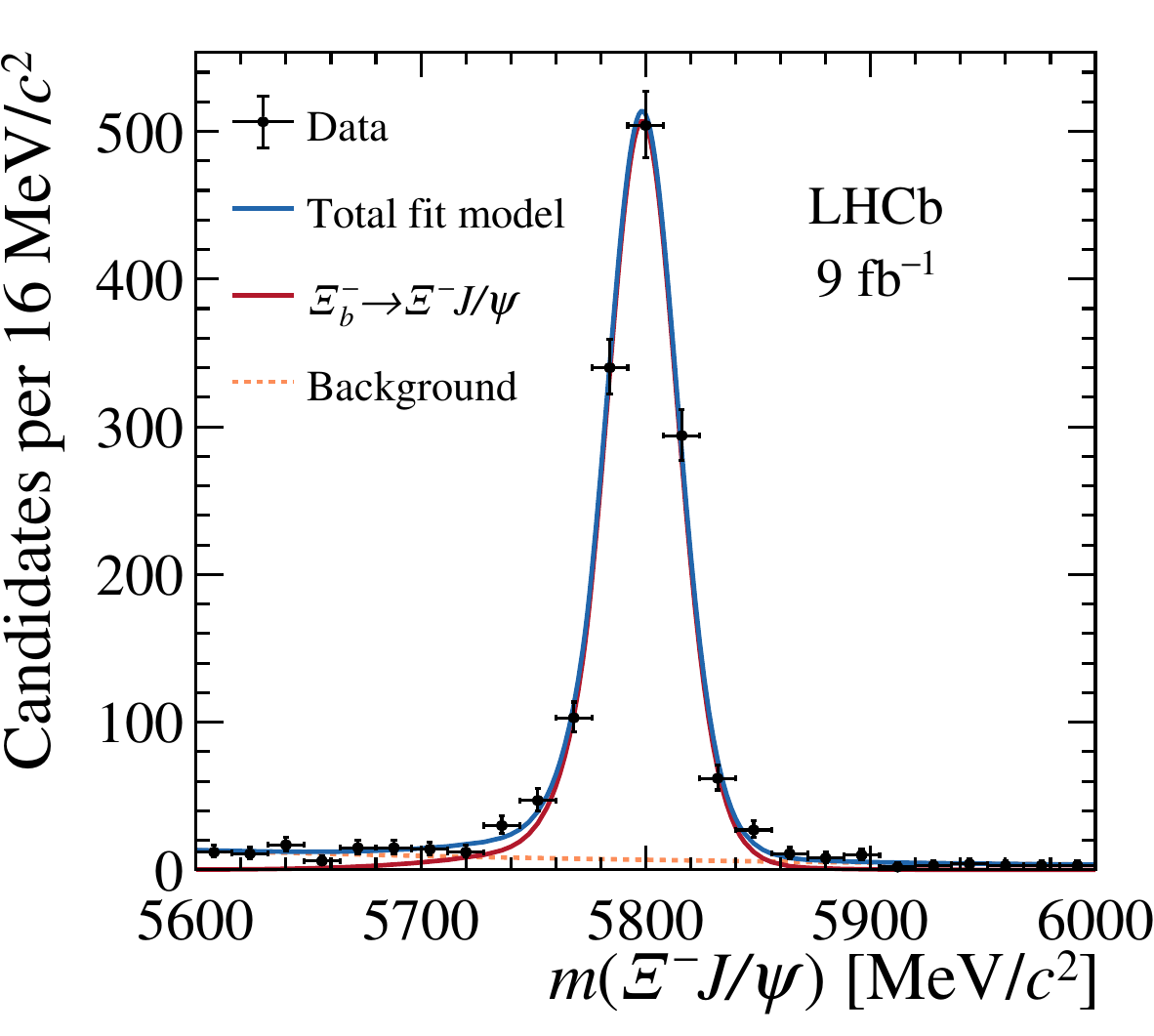}
    \end{subfigure}
    \caption{Invariant-mass distributions in the (left) \XibToXimm dataset in the full nonresonant dimuon invariant-mass range and (right) \XibToXiJPsi dataset, all data-taking periods combined. The results of the simultaneous fit are overlaid, with the (full red line) signal  and (dotted line) background  components also shown.}
    \label{fig:Ximm_obs}
\end{figure}

In the search for the \XibToXimm decay mode, candidates are considered in the entire dimuon invariant-mass range excluding the regions (8.4, 11.0)\gevgevcccc and (12.5, 14.0)\gevgevcccc that are contaminated by the charmonia resonances and their resolution tails. 
A simultaneous fit is performed to the \XibToXimm candidates and the normalisation \XibToXiJPsi candidates. All data-taking periods are combined, due to the small size of the former decay sample. 
In this fit, the $\Xim\mumu$ invariant mass is calculated by constraining the invariant mass of the \Xim and \Lz hyperons to their known values, and constraining the \Xibm candidate to originate from the PV. No mass constraint is applied for the dimuon pair, leading to a worse invariant-mass resolution compared to that in Sec.~\ref{sec:BFpsi}. 
The fit model is equivalent to that for the \rbrpsi measurement.
However in the \XibToXiJPsi case, an additional Gaussian component is added to the two double-sided Crystal Ball functions in order to improve the description of data. 
The free parameters of the signal in the fit to data are the yields of both decay modes, and the \Xibm mass for both data-taking periods which is common between the two decay channels. 
The background is described in each dataset by an exponential function, with a freely varying normalisation and slope. The corresponding datasets and the fit result are shown in Fig.~\ref{fig:Ximm_obs}. 
The yield of the \XibToXimm decay is $70 \pm 10$. The profile log-likelihood scan for this observable is shown in App.~\ref{app:pll}.
Using Wilks' theorem~\cite{Wilks:1938dza}, the statistical significance of the signal is estimated to be about 9 standard deviations. This constitutes the first observation of \XibToXimm decay. 
The resulting yield of the \XibToXiJPsi decay is $1373 \pm 39$, with the difference compared to the yield reported in Sec.~\ref{sec:BFpsi} being due to the tighter BDT requirement.

\begin{figure}[t]
    \centering
        \includegraphics[height=2.7in]{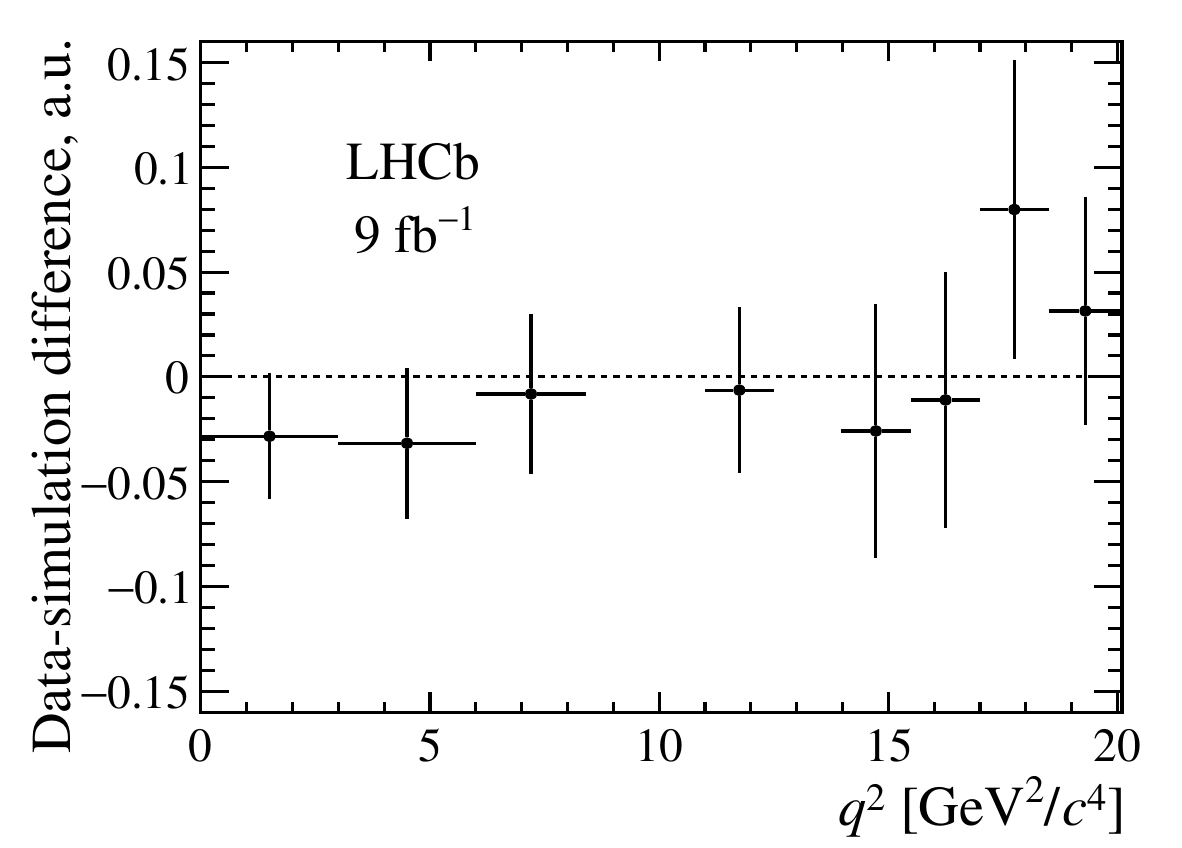}
    \caption{Absolute difference between the normalised \qsq distributions in data and corrected simulation: no significant discrepancies are observed. No efficiency correction is applied. The uncertainties are statistical only.}
    \label{fig:Ximm_q2_pull}
\end{figure}

Checks are performed to validate whether the corrected simulation models the data sufficiently well. For that, the background is subtracted using the \sPlot procedure~\cite{Pivk:2004ty}.
A satisfactory agreement is observed  within statistical uncertainties for the \qsq distribution, decay-angle distributions, and variables used in the signal selection. Individual invariant-mass fits performed in narrow intervals of \qsq also show a good agreement between data and simulation in the normalised \qsq distributions, as shown in Fig.~\ref{fig:Ximm_q2_pull}. 

\section{Measurement of the \XibToXimm decay rate}
\label{sec:BFrare}

A simultaneous fit is performed to the \XibToXiJPsi and \XibToXimm candidates, with each of the three \qsq regions considered as a separate fit projection. Good agreement between the Run 1 and Run 2 results is observed for the \rbrpsi ratio. The two datasets for the statistically limited measurement of the rare \XibToXimm decay are thus merged in the following.
The signal and background models are similar to those described in Sec.~\ref{sec:Observation}. 
The free parameters of the fit are the branching fraction ratios in each of the three \qsq regions
\begin{equation}
 \begin{split}
 & r_{\mathrm{BR}}^{i} \equiv \frac{\BR(\XibToXimm)_{i}}{\BR(\XibToXiJPsi)} = 
 \\
 &\frac{N(\XibToXimm)_i}{N(\XibToXiJPsi)} \times \frac{\varepsilon(\XibToXiJPsi)}{\varepsilon(\XibToXimm)_i} \times {\BR(\jpsi\to\mumu)},
\end{split}
\label{eq:r_br}
\end{equation}
with $i=$ I, II, III 
together with the yield of the \XibToXiJPsi signal, the respective signal and background yields and slopes of combinatorial background in each of the four fit projections, and the value of the \Xibm mass, shared between the four fit projections.

\begin{figure}[t]
    \centering

    \begin{subfigure}[t]{0.5\textwidth}
        \centering
        \includegraphics[height=2.8in]{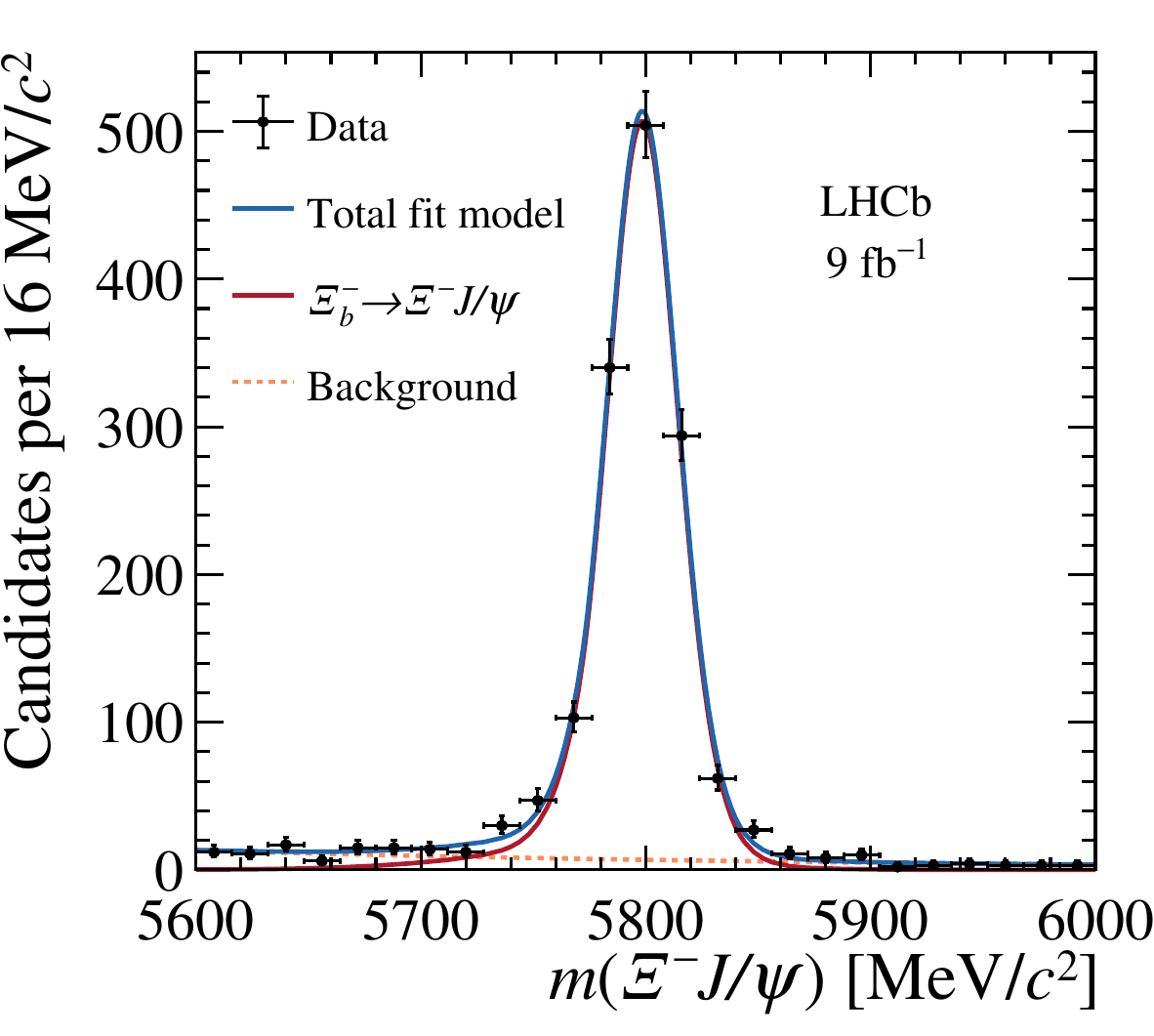}
    \end{subfigure}%
    \begin{subfigure}[t]{0.5\textwidth}
        \centering
        \includegraphics[height=2.8in]{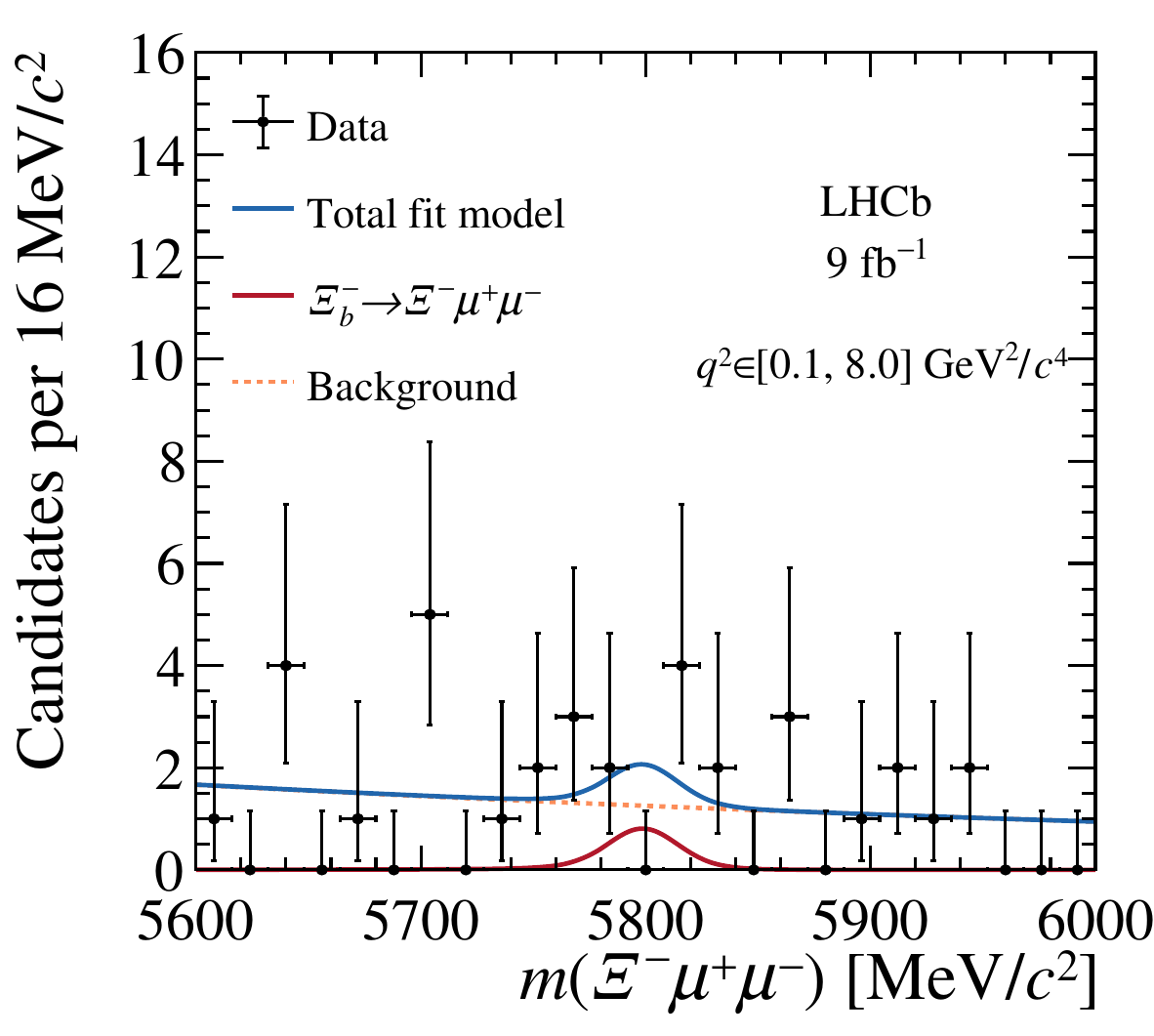}
        \label{fig:blind_fit_Ximm_wsyst_ALT_I}
    \end{subfigure}

    \begin{subfigure}[t]{0.5\textwidth}
        \centering
        \includegraphics[height=2.8in]{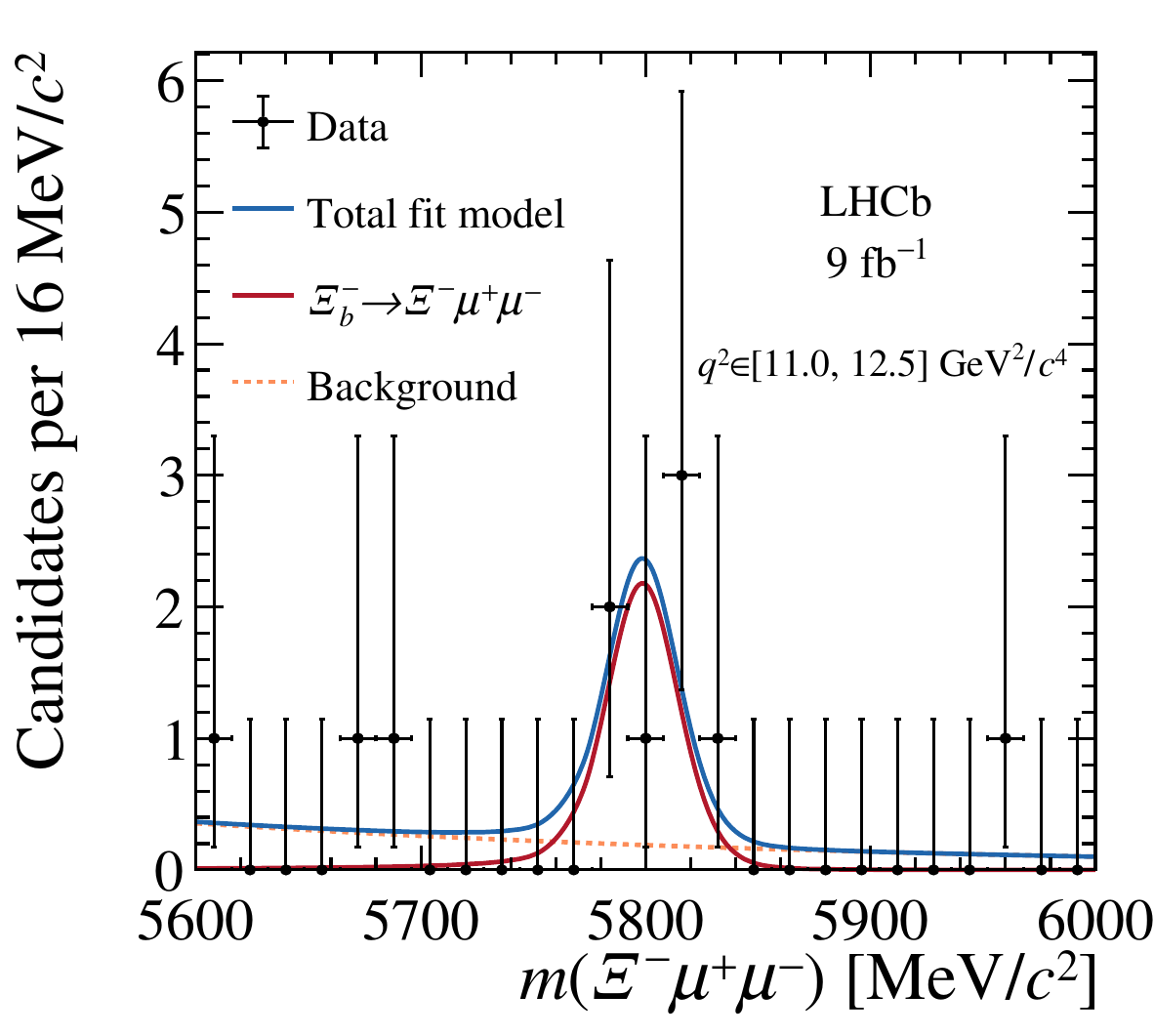}
        \label{fig:blind_fit_Ximm_wsyst_ALT_II}
    \end{subfigure}%
    \begin{subfigure}[t]{0.5\textwidth}
        \centering
        \includegraphics[height=2.8in]{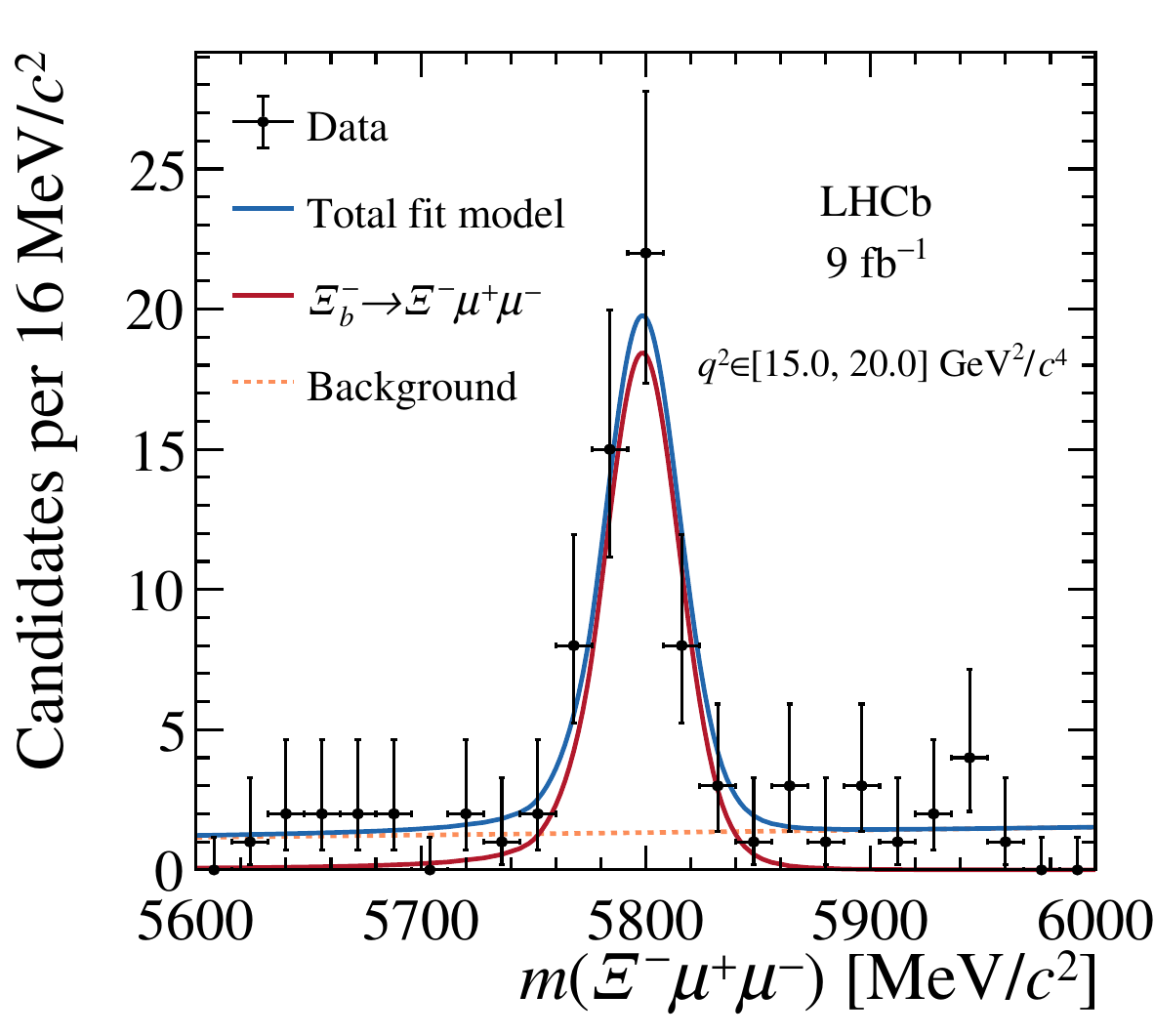}
        \label{fig:blind_fit_Ximm_wsyst_ALT_III}
    \end{subfigure}

    \caption{Invariant-mass distributions in the \XibToXiJPsi and \XibToXimm datasets. The results of the simultaneous fit are overlaid, with the (full red line) signal and (dotted line) background components also shown.}
    \label{fig:blind_fit_Ximm_wsyst_ALT}
\end{figure}

The fit projections are shown in Fig.~\ref{fig:blind_fit_Ximm_wsyst_ALT}. The \XibToXiJPsi dataset is identical to that in Fig.~\ref{fig:Ximm_obs}.  
In the three regions considered, the statistical significance of signal, calculated using Wilks’ theorem~\cite{Wilks:1938dza}, is found to be 0.6, 3.1 and 10 standard deviations. 
In \qsq region II, pseudoexperiments show that the probability of
background processes producing a mass distribution that is less compatible with the
background-only model than the data, including both statistical and systematic uncertainties, is $1.3 \times 10^{-3}$, corresponding to a significance of 3.0 standard deviations.

Possible sources of systematic uncertainty are studied in detail and summarised in Table~\ref{tab:syst_rbr} for \qsq regions I, II and III separately. 
The size of the simulated samples, used to estimate the signal efficiencies, leads to a $0.6-0.8\%$ relative uncertainty on the efficiency ratio. The procedures used to estimate the uncertainties related to the correction of the simulation, muon identification performance, downstream tracking efficiency, knowledge of the \Xibm lifetime, and hyperon material interactions, are repeated in the same manner as for the \rbrpsi measurement. The uncertainty related to the correction of simulation ranges from 1.5\% in the \qsq region~II 
to 3.5\% in the \qsq region I, which is the broadest, covering a vast kinematic range of \Xibm decay products. 

As mentioned in Sec.~\ref{sec:Observation}, the agreement of the \qsq and angular distributions between data and simulation is found to be satisfactory within statistical uncertainties. The effect of the angular weighting is tested by replacing it with the momentum-asymmetry weighting used in the resonant channels, leading to~$1-3\%$ uncertainty. 

Several sources of systematic uncertainty related to the fit model are considered. The uncertainties related to fit bias and the chosen signal and background models are studied using pseudoexperiments. The possible background from \LbToLmm (\LbToLJPsi) decays combined with a random pion is examined. In the same way as for the \rbrpsi measurement, the flavour SU(3) symmetry is invoked to estimate the expected yields of the \Lb background and generate pseudoexperiments. The yields of the \LbToLmm decays are however allowed to fluctuate within a 30\% uncertainty to account for possible SU(3)-breaking corrections. 
The relative impact of the possible \Lb background on the measurement of \rbr is found to be at the level of $1.9-10.4\%$. 
The final relative systematic uncertainty due to the fit model is $3.0 - 14.3\%$. 
Lastly, the uncertainties on the branching fractions of the dimuon decay of the \jpsi meson are propagated to the final result, leading to a relative uncertainty of $0.6\%$. 
\begin{table}[t]
  \centering
  \caption{Relative systematic uncertainties on $r_{\mathrm{BR}}^{i}$. The last column specifies whether the values can be considered fully correlated between the \qsq regions or fully uncorrelated.}
  \label{tab:syst_rbr}
  \begin{tabular}{c|c|c|c|c}
   Type & Region I [\%] &  Region II [\%] & Region III [\%] & Correlated\\
  \hline
  Size of simulated samples &  \phantom{1}0.6 & \phantom{1}0.8 & 0.7 & no \\
  Correction of simulation &  \phantom{1}3.5 & \phantom{1}1.5 & 2.3 & yes\\
  Downstream tracking &  \phantom{1}1.6 & \phantom{1}1.6 & 1.6 & yes \\
  Material interactions &  \phantom{1}1.1 & \phantom{1}1.1 & 1.1 & yes \\
  Decay model &  \phantom{1}1.0 & \phantom{1}2.3 & 3.0 & no \\ 
  Fit model &  14.3 & 11.8 & 3.0 & no \\
  External input ($\mathcal{B}$) &  \phantom{1}0.6 & \phantom{1}0.6 & 0.6 & yes\\
       \hline 
 {Total} & {15.1} & {12.4} & {5.2} & \\ 
 \end{tabular}
\end{table}

\begin{figure}[t]
    \centering
   \begin{subfigure}[t]{1\textwidth}
        \centering
        \includegraphics[height=2.8in]{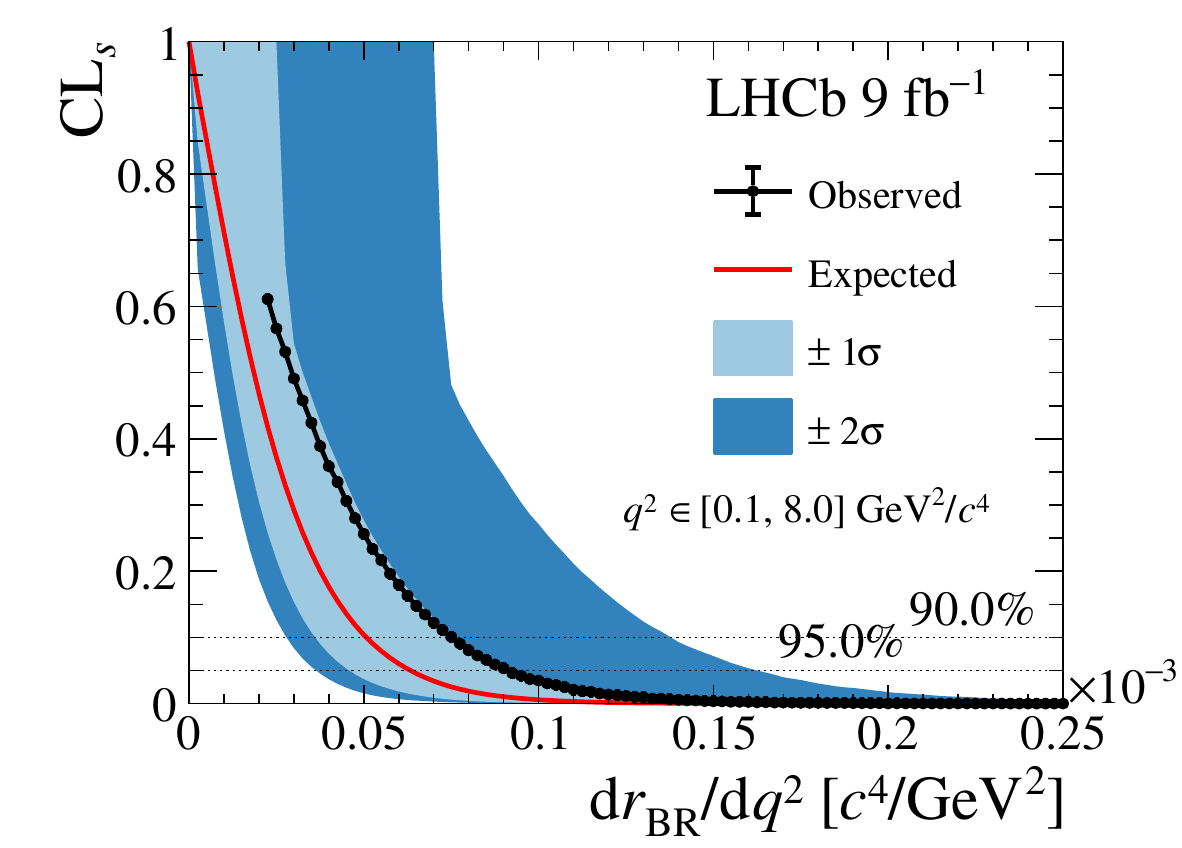}
        \end{subfigure}%
    \caption{Result of the $\mathrm{CL}_s$ scans in the background-only hypothesis for the d$\rbr/$d$\qsq$ observable in region I. The red line describes the background-only expectation, while the $1\sigma$ and $2\sigma$ bands are shown as light and dark blue bands, respectively. The black dotted line corresponds to the observed limits. The two dashed lines intersecting with the latter indicate the limits at 90\% and 95\% CL for the upper and lower line, respectively.
    }
    \label{fig:gammacombo_expected_limits}
\end{figure}

\begin{table}[t]
  \centering
  \caption{Values of d$\rbr/$d$\qsq$ obtained in the three \qsq regions under study.}
  \label{tab:diff_rbr}
  \begin{tabular}{l|c}
    \qsq region (\gevgevcccc) & d$\rbr/$d$\qsq$ $[\times 10^{-4} (\!\gevgevcccc)^{-1}]$\\
    \hline
    $\qsq \in [\phantom{1}0.1, \phantom{1}8.0]$ & $0.19 \pm 0.33\stat \pm 0.03\syst$\rule{0pt}{0.9\normalbaselineskip}\\
    $\qsq \in [11.0,12.5]$ & $1.47 \pm 0.67\stat \pm 0.20 \syst$\\ 
    $\qsq \in [15.0,20.0]$ & $2.46 \pm 0.38 \stat \pm 0.12 \syst$ \\
 \end{tabular}
\end{table}
The results are summarised in Table~\ref{tab:diff_rbr}, where the $d\rbr/d\qsq$ is quoted. The profile log-likelihood scans for the d$\rbr/$d$\qsq$ observable in the three \qsq intervals are shown in App.~\ref{app:pll}.
While a significant signal is observed in \qsq region III and evidence for signal is seen in region II, the signal significance does not reach the threshold of evidence in the \qsq region I. Upper limits
are set on the d$\rbr/$d$\qsq$ in the \qsq region I using the $\text{CL}_{s}$ method~\cite{CLs} as implemented in the \textsc{GammaCombo} package~\cite{GammaCombo,LHCb-PAPER-2016-032} using a one-sided test statistic.
The corresponding $\mathrm{CL}_s$ scans are shown in Fig.~\ref{fig:gammacombo_expected_limits}. The observed (expected under background-only hypothesis) upper limit is $0.75$ (0.49) at 90\% CL, or $0.91$ (0.62) at 95\% CL, in units of $10^{-4}$ $c^4/\gev^2$.

\subsection{Absolute decay rate}

In order to facilitate comparison with the theoretical expectations, the absolute branching fraction of the \XibToXimm decay is estimated as follows. 

Having previously determined the branching fraction of the \XibToXimm  decay relative to the \XibToXiJPsi decay, one requires the knowledge of the branching fraction of the normalisation \XibToXiJPsi decay in order to obtain its absolute value.
The only existing measurements, performed by the CDF~\cite{Aaltonen:2009ny} and LHCb~\cite{LHCb-PAPER-2018-047} experiments, are not absolute measurements of the ratio of branching fractions, but rather of the quantity
\begin{equation}
    \frac{f_\Xibm}{f_\Lb}\frac{\BR(\XibToXiJPsi)}{\BR(\LbToLJPsi)},
\end{equation}
where $f$ denotes the hadronisation fraction of a \bquark quark into the corresponding baryon. 
Since no independent measurements exist for the ${f_\Xibm}/{f_\Lb}$ ratio, these measurements cannot be used to infer the value of $\BR(\XibToXiJPsi)$. The SU(3) relation~\cite{Savage:1989jx,Voloshin:2015xxa,Hsiao:2015txa} is exploited instead:
\begin{equation}
    \frac{\Gamma(\XibToXiJPsi)}{\Gamma(\LbToLJPsi)} = \frac{3}{2},
    \label{eq:Gsu3f}
\end{equation}
where $\Gamma$ is the partial decay width. Using the latest LHCb measurement of the ratio of lifetimes of the \Xibm and \Lb baryons, $1.078\pm0.014$~\cite{LHCb-PAPER-2024-010}, one obtains
\begin{equation}
    \frac{\BR(\XibToXiJPsi)}{\BR(\LbToLJPsi)} = \frac{3}{2} \times 1.078 \approx 1.6;
    \label{eq:su3f}
\end{equation}
which is usually~\cite{LHCb-PAPER-2018-047} taken with an uncertainty of 30\%, the typical size of SU(3)-breaking effects. An additional uncertainty of 1.3\% arises from the precision on the ratio of the lifetimes. 

The value of $\BR(\LbToLJPsi)$ has recently been measured by LHCb~\cite{LHCb-PAPER-2025-035} to be \mbox{$(3.34 \pm 0.31) \times 10^{-4}$}.
This gives
\begin{equation*}
    \BR(\XibToXiJPsi) = (5.3\pm0.5\pm1.6 \pm 0.1) \times 10^{-4} = (5.3\pm1.7) \times 10^{-4},
    \label{eq:BR_XibToXiJPsi}
\end{equation*}
where the first uncertainty arises from that on $\BR(\LbToLJPsi)$, the second from SU(3) breaking, and the third from the lifetime uncertainties; the last expression combines all uncertainties in quadrature.

The differential values, d$\BR(\XibToXimm)/$d$\qsq$, obtained in the studied dimuon invariant-mass regions are listed in Table~\ref{tab:diff_BF}. In the \qsq region I where no significant signal is observed, an upper limit is estimated, taking into account the additional uncertainty arising from $\BR(\XibToXiJPsi)$. The observed (expected under background-only hypothesis) upper limit is $4.7$ (3.0) at 90\% CL, or $6.4$ (4.0) at 95\% CL, in units of $10^{-8}(\!\gevgevcccc)^{-1}$.
\begin{table}[b]
  \centering
  \caption{Differential branching fraction of \XibToXimm decay obtained in the different \qsq regions.} 
  \label{tab:diff_BF}
  \begin{tabular}{l|c}
    \qsq region (\gevgevcccc)& d$\BR(\XibToXimm)/$d$\qsq$ $[\times 10^{-8} (\!\gevgevcccc)^{-1} ]$\\
    \hline
    $\qsq \in [0.1, 8.0]$\rule{0pt}{0.9\normalbaselineskip} 
    & $\phantom{1}1.0 \pm 1.8 \stat \pm 0.2 \syst\pm 0.3\,(\textrm{norm})$\\
    $\qsq \in [11.0, 12.5]$ 
    & $\phantom{1}7.8 \pm 3.5 \stat \pm 1.0 \syst\pm 2.5\,(\textrm{norm})$\\
    $\qsq \in [15.0, 20.0]$ & $13.0 \pm 2.0\, (\textrm{stat}) \pm 0.6 \syst \pm 4.2 \,(\textrm{norm})$\\
 \end{tabular}
\end{table}

\begin{figure}[tb]
    \centering
        \includegraphics[height=2.8in]{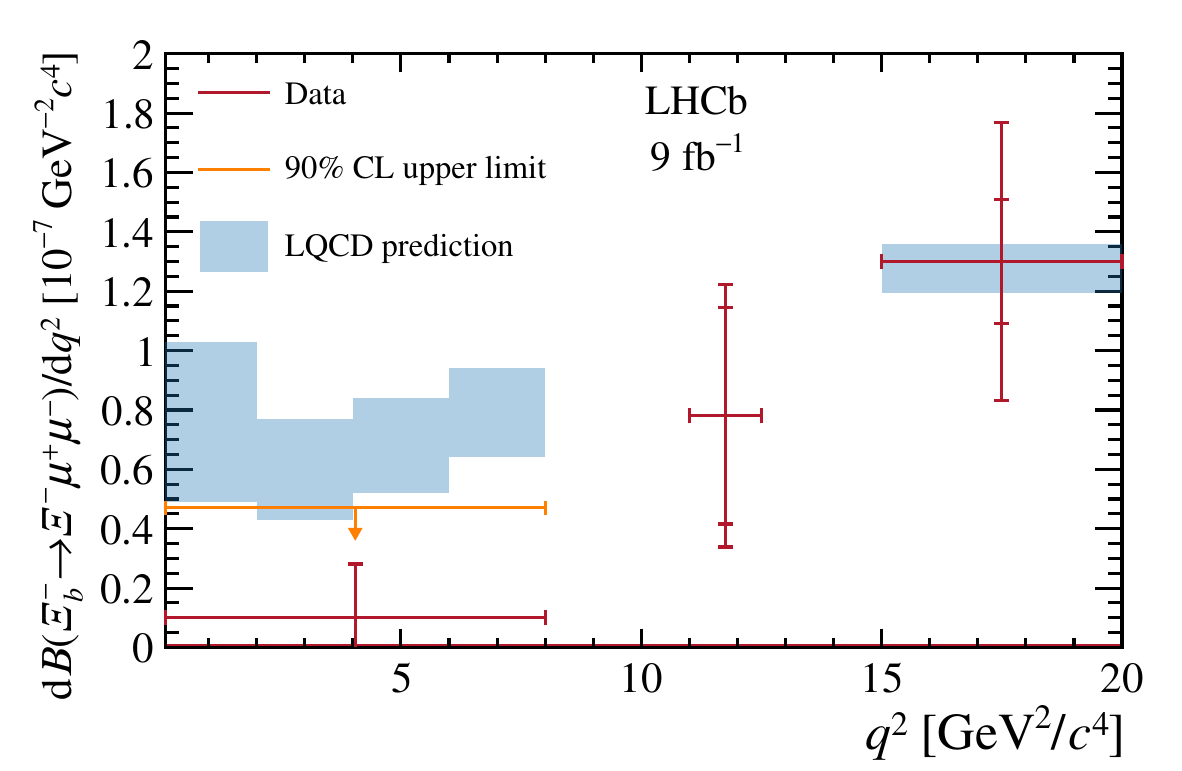}
    \caption{Differential branching fraction d$\BR/$d$\qsq$ as a function of \qsq. Shown are the (dark red markers) obtained values with the inner error bars showing the statistical uncertainties and the outer ones showing the total uncertainties. The upper limits at 90\% CL (orange markers), and the (filled blue areas) lattice QCD prediction~\cite{Farrell:2026swf}, are also shown.} 
    \label{fig:results_vs_SM}
\end{figure}

The values of d$\BR(\XibToXimm)/$d$\qsq$ are shown in Fig.~\ref{fig:results_vs_SM} as a function of~\qsq and compared to the lattice QCD prediction from Ref.~\cite{Farrell:2026swf}. In region~III, where a direct comparison can be made, a good agreement is observed. In  region~I, the 95\% CL upper limit is comparable to the SM value. 

The values obtained in the three \qsq regions are summed, 
accounting for the vetoed charmonium regions. Using the corrected simulated \XibToXimm dataset, the fraction of \XibToXimm candidates in the considered \qsq regions is determined to be about 72\%. 
This gives the integrated value  

\begin{equation*}
    \BR(\XibToXimm) = { (118 \pm 24 \stat \pm 7 \syst \pm 38 \,(\textrm{norm})) \times 10^{-8}}.
\end{equation*} 

If compared to the experimental value of $\BR(\LbToLmm)$~\cite{LHCb-PAPER-2015-009} in the region $\qsq \in [15.0, 20.0] \gevgevcccc$, corrected for the latest knowledge of $\BR(\LbToLJPsi)$~\cite{LHCb-PAPER-2025-035,Bordone:2025elp}, which gives d$\BR/$d$\qsq (\LbToLmm) = (6.4 \pm 0.8) \times 10^{-8} (\!\gevgevcccc)^{-1}$ , the rate of the \XibToXimm decay is enhanced compared to the \LbToLmm decay, as predicted by the SU(3) symmetry.

\section{Conclusion}
\label{sec:Conclusion}

The rare decay \XibToXimm is observed for the first time using the combined LHCb datasets collected in Run 1 and 2, with signal significance greatly exceeding 5 standard deviations. The decay is studied in bins of the dimuon invariant-mass squared, \qsq. Significant signal is observed in the interval $\qsq \in [15.0, 20.0]\gevgevcccc$, evidence of signal is seen in the range $\qsq \in [11.0, 12.5\gevgevcccc]$, while the signal significance remains low in the region $\qsq \in [0.1, 8.0]\gevgevcccc$. 
The differential branching fraction of this decay is obtained and upper limits are reported for the \qsq regions where the signal significance is low. 
The measured differential branching fraction at the high-\qsq, \mbox{$(13.0 \pm 2.0 \stat \pm 0.6 \syst \pm 4.2 \,(\textrm{norm})) \times 10^{-8} (\!\gevgevcccc)^{-1}$}, is consistent with the recent lattice QCD prediction~\cite{Farrell:2026swf}. Similarly, the low-\qsq upper limit is consistent with the lower range of the lattice QCD prediction. The integrated branching fraction is measured to be ${ (118 \pm 24 \stat \pm 7 \syst \pm 38 \,(\textrm{norm})) \times 10^{-8}}$. 
It is noteworthy that the measurement suggests that the branching fraction of the \XibToXimm is one of the highest among all observed \bsll transitions. 

The decay \XibToXiPsi is also studied and the ratio of branching fractions of \XibToXiPsi and \XibToXiJPsi decays is measured to be ${  0.491 \pm 0.041 \stat \pm 0.019 \syst}$. This result is lower than, but consistent with, the value measured previously by the CMS collaboration~\cite{CMS:2024rbi}, and improves the relative precision on it by a factor of 3. It is in good agreement with the equivalent ratio measured in \Lb~\cite{Aad:2015msa,LHCb-PAPER-2017-041} and \Xibz decays~\cite{LHCb-PAPER-2026-030}. 

The observation of the decay \XibToXimm opens opportunities to study its rich angular structure with the larger dataset collected by LHCb in Run 3. Together with the observation of the $\Xibz \to \Xiz \jpsi$ decay~\cite{LHCb-PAPER-2026-030}, it lays the groundwork for future studies of the rare decays of \Xib and \Omegab baryons.

%% file: LHCb/acknowledgements.tex
\section*{Acknowledgements}
%
%
\noindent We express our gratitude to our colleagues in the CERN
accelerator departments for the excellent performance of the LHC. We
thank the technical and administrative staff at the LHCb
institutes.
We acknowledge support from CERN and from the national agencies:
ARC (Australia);
CAPES, CNPq, FAPERJ and FINEP (Brazil); 
MOST and NSFC (China); 
CNRS/IN2P3 and CEA (France);  
BMFTR, DFG and MPG (Germany);
NKFIH (Hungary);              
INFN (Italy); 
NWO (Netherlands); 
MNiSW and NCN (Poland); 
MEC/IFA (Romania); 
MICIU and AEI (Spain);
SNSF and SER (Switzerland); 
NASU (Ukraine); 
STFC (United Kingdom); 
DOE NP and NSF (USA).
We acknowledge the computing resources that are provided by ARDC (Australia), 
CBPF (Brazil),
CERN, 
IHEP and LZU (China),
IN2P3 (France), 
KIT and DESY (Germany), 
INFN (Italy), 
SURF (Netherlands),
Polish WLCG (Poland),
IFIN-HH (Romania), 
PIC (Spain), CSCS (Switzerland), 
GridPP (United Kingdom),
and NSF (USA).  
We are indebted to the communities behind the multiple open-source
software packages on which we depend.
Individual groups or members have received support from
RTP (Australia), 
FWO Odysseus grant G0ASD25N (Belgium), 
Key Research Program of Frontier Sciences of CAS, CAS PIFI, CAS CCEPP (China); 
Minciencias (Colombia);
EPLANET, Marie Sk\l{}odowska-Curie Actions, ERC and NextGenerationEU (European Union);
A*MIDEX, ANR, IPhU and Labex P2IO, and R\'{e}gion Auvergne-Rh\^{o}ne-Alpes (France);
Alexander-von-Humboldt Foundation (Germany);
ICSC (Italy); 
Severo Ochoa and Mar\'ia de Maeztu Units of Excellence, GVA, XuntaGal, GENCAT, InTalent-Inditex and Prog.~Atracci\'on Talento CM (Spain);
the Leverhulme Trust, the Royal Society and UKRI (United Kingdom).

%% file: appendix.tex

\newpage
\section*{Appendix}

\appendix

\section{Profile log-likelihood scans}
\label{app:pll}
\begin{figure}[b]
    \centering
    \begin{subfigure}[t]{0.48\textwidth}
        \centering
        \includegraphics[width=.95\textwidth]{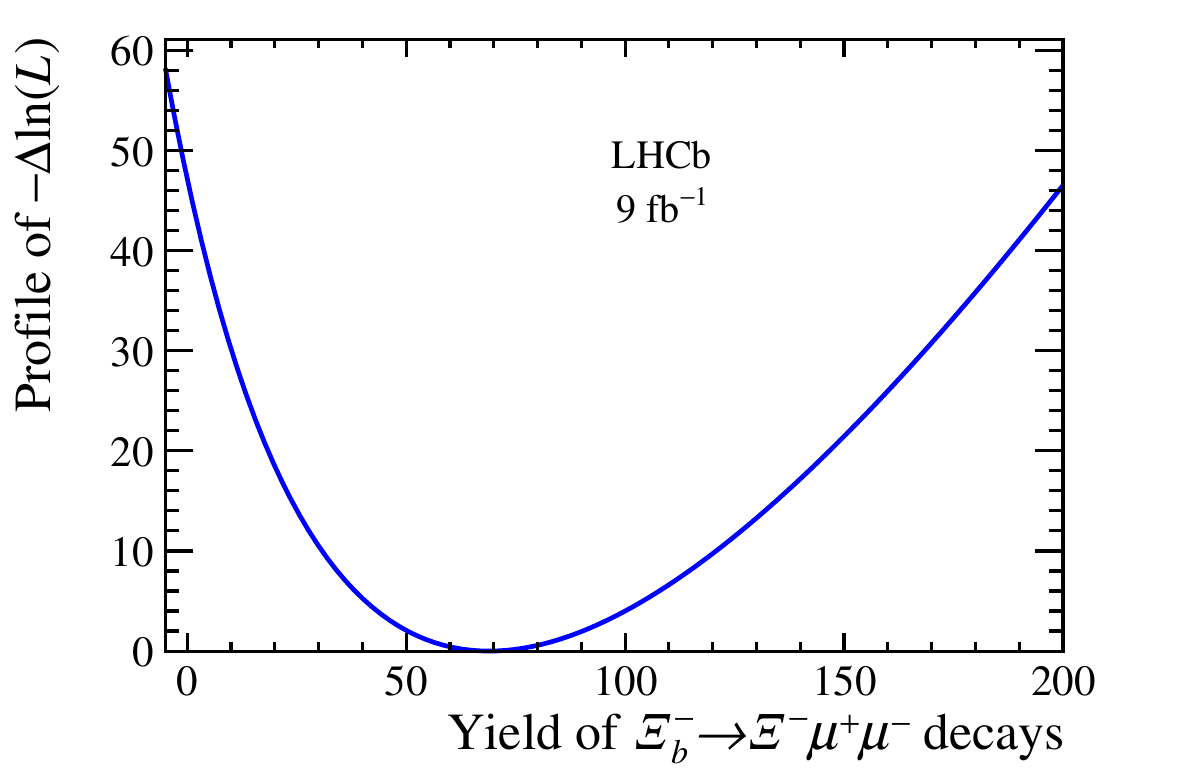}
        \caption{Yield of the \XibToXimm decays, corresponding to the fit shown in Fig.~\ref{fig:Ximm_obs}.}
    \end{subfigure}\hfill
     \begin{subfigure}[t]{0.48\textwidth}
        \centering
        \includegraphics[width=.95\textwidth]{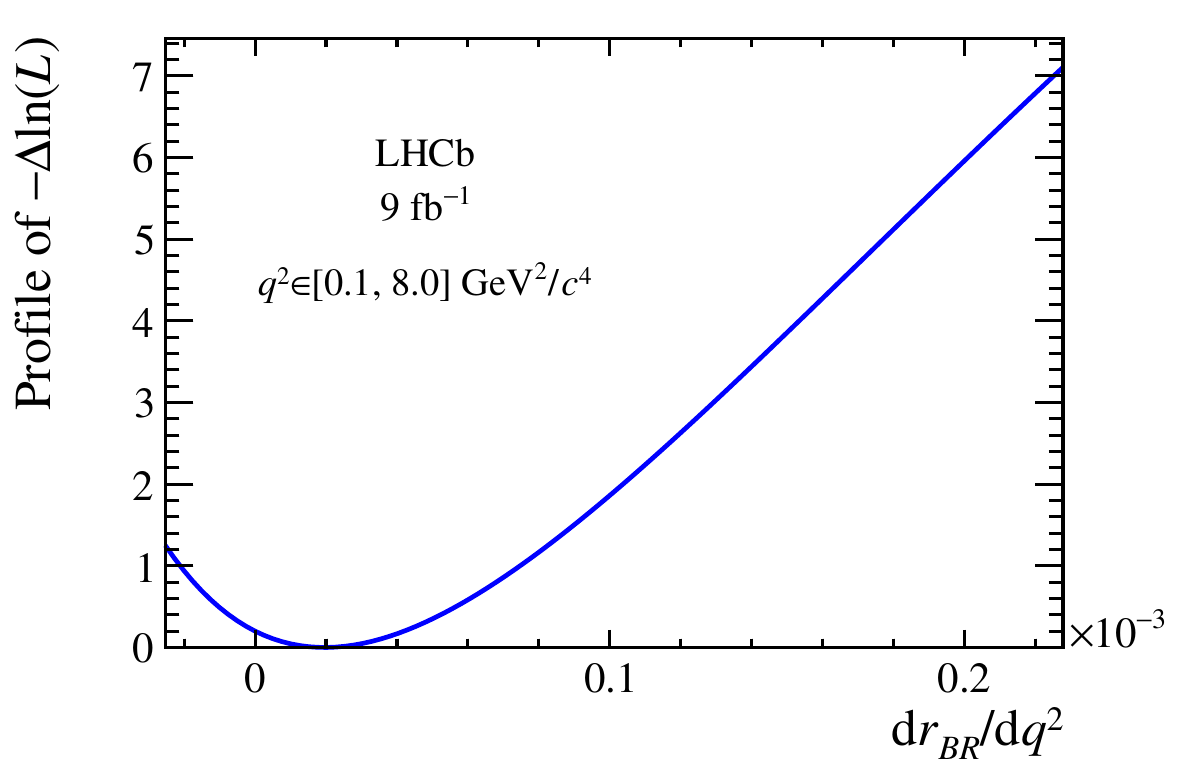}
        \caption{\rbr in the \qsq region I, corresponding to the fit shown in Fig.~\ref{fig:blind_fit_Ximm_wsyst_ALT}.}
    \end{subfigure}

     \begin{subfigure}[t]{0.48\textwidth}
        \centering
        \includegraphics[width=.95\textwidth]{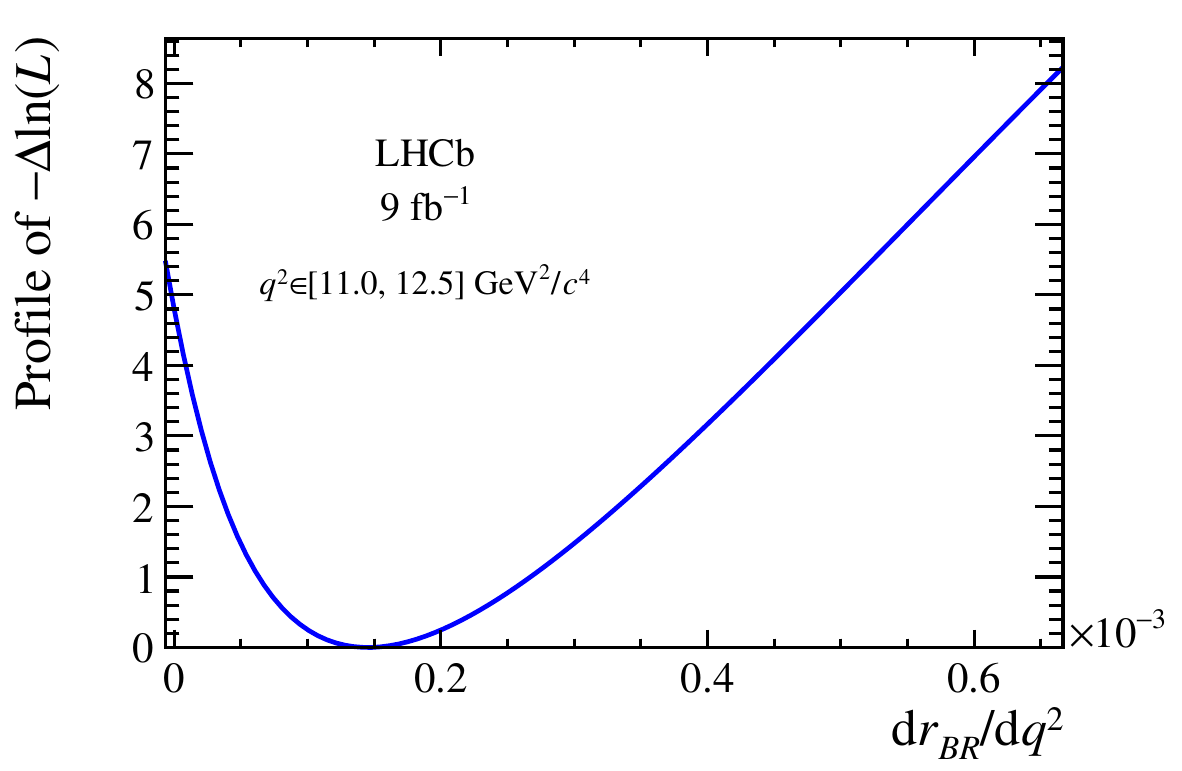}
        \caption{\rbr in the \qsq region II, corresponding to the fit shown in Fig.~\ref{fig:blind_fit_Ximm_wsyst_ALT}.}
    \end{subfigure}\hfill
     \begin{subfigure}[t]{0.48\textwidth}
        \centering
        \includegraphics[width=.95\textwidth]{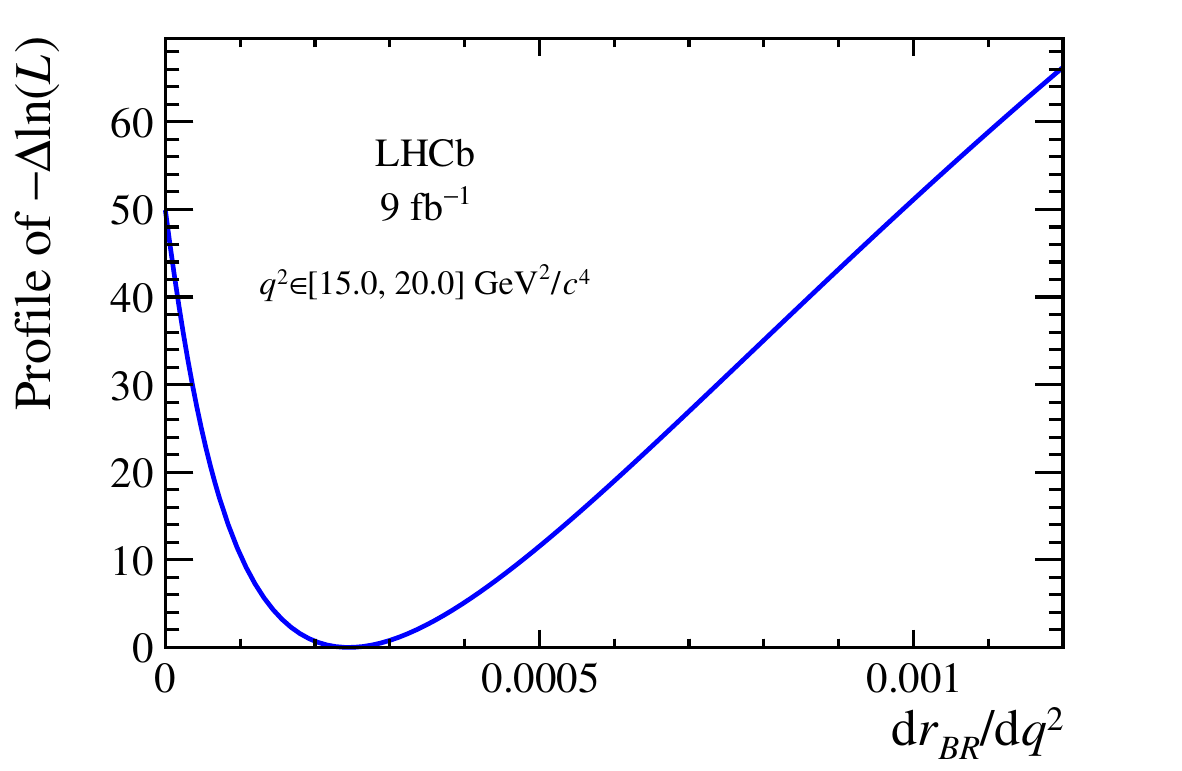}
        \caption{\rbr in the \qsq region III, corresponding to the fit shown in Fig.~\ref{fig:blind_fit_Ximm_wsyst_ALT}.}
    \end{subfigure}

     \begin{subfigure}[t]{0.48\textwidth}
        \centering
        \includegraphics[width=.95\textwidth]{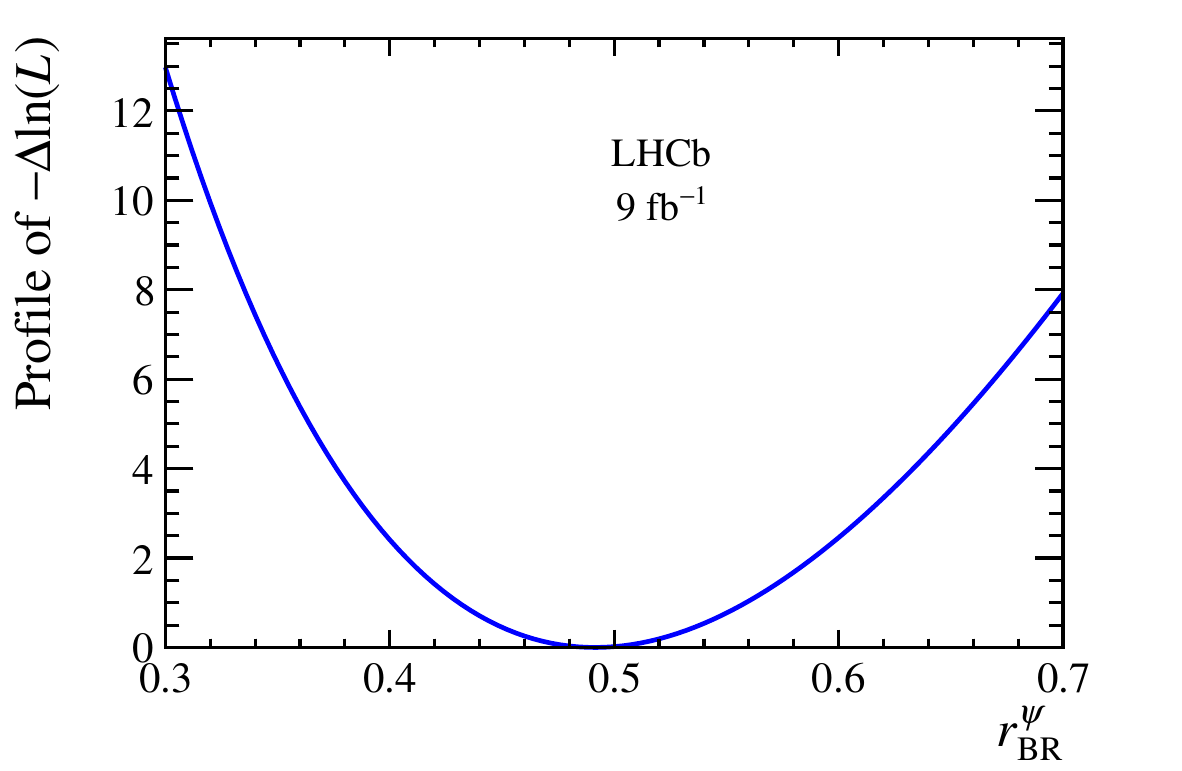}
        \caption{\rbrpsi, corresponding to the fit shown in Fig.~\ref{fig:simfit_psi2s_wsyst}.}
    \end{subfigure}

    \caption{Profile log-likelihood scans for the different observables measured in this paper. }
    \label{fig:PLL}
\end{figure}

Profile log-likelihood scans for the key observables measured in this paper are shown in Fig.~\ref{fig:PLL}. 
\clearpage

%% file: LHCb/Authorship_LHCb-PAPER-2026-036.tex
\centerline
{\large\bf LHCb collaboration}
\begin
{flushleft}
\small
R.~Aaij$^{41}$\lhcborcid{0000-0003-0533-1952},
M.~Abdelfatah$^{73}$,
A.S.W.~Abdelmotteleb$^{61}$\lhcborcid{0000-0001-7905-0542},
C.~Abellan~Beteta$^{55}$\lhcborcid{0009-0009-0869-6798},
F.~Abudin\'en$^{63}$\lhcborcid{0000-0002-6737-3528},
T.~Ackernley$^{65}$\lhcborcid{0000-0002-5951-3498},
A.A.~Adefisoye$^{73}$\lhcborcid{0000-0003-2448-1550},
B.~Adeva$^{51}$\lhcborcid{0000-0001-9756-3712},
M.~Adinolfi$^{59}$\lhcborcid{0000-0002-1326-1264},
P.~Adlarson$^{89,46}$\lhcborcid{0000-0001-6280-3851}
% ...

{\footnotesize \it

% ...
$^{41}$Nikhef National Institute for Subatomic Physics, Amsterdam, Netherlands\\
% ...
$^{46}$National Center for Nuclear Research (NCBJ), Warsaw, Poland\\
% ...
$^{51}$Instituto Galego de F{\'\i}sica de Altas Enerx{\'\i}as (IGFAE), Universidade de Santiago de Compostela, Santiago de Compostela, Spain\\
% ...
$^{55}$Physik-Institut, Universit{\"a}t Z{\"u}rich, Z{\"u}rich, Switzerland\\
% ...
$^{59}$H.H. Wills Physics Laboratory, University of Bristol, Bristol, United Kingdom\\
% ...
$^{61}$Department of Physics, University of Warwick, Coventry, United Kingdom\\
% ...
$^{63}$School of Physics and Astronomy, University of Edinburgh, Edinburgh, United Kingdom\\
$^{64}$School of Physics and Astronomy, University of Glasgow, Glasgow, United Kingdom\\
$^{65}$Oliver Lodge Laboratory, University of Liverpool, Liverpool, United Kingdom\\
% ...
$^{73}$Syracuse University, Syracuse, NY, United States\\
% ...
$^{89}$Department of Physics and Astronomy, Uppsala University, Uppsala, Sweden, associated to $^{64}$\\
% ...
}
\end{flushleft}